\documentclass[11pt]{article}  
\usepackage{amsmath,amssymb,graphicx,color}

\newtheorem{theorem}{Theorem}[section]
\newtheorem{lemma}[theorem]{Lemma}
\newtheorem{claim}[theorem]{Claim}

\newtheorem{corollary}[theorem]{Corollary}

\begin{document}

\def\reals{{\mathbb R}}
\def\sph{{\mathbb S}}
\def\eps{{\varepsilon}}
\def\bd{{\partial}}
\def\A{{\cal A}}
\def\AV{{\cal A}^\bot}
\def\B{{\cal B}}
\def\C{{\cal C}}
\def\D{{\cal D}}
\def\EE{{\mathbb E}}
\def\F{{\cal F}}
\def\H{{\cal H}}
\def\K{{\cal K}}
\def\Q{{\cal Q}}
\def\R{{\cal R}}
\def\S{{\cal S}}
\def\V{{\cal V}}
\def\T{{\cal T}}
\def\vv{{\bf v}}
\def\V1{{\cal V}_1}
\def\Z{{\cal Z}}

\def\marrow{{\marginpar[\hfill$\longrightarrow$]{$\longleftarrow$}}}

\def\hayim#1{{\sc Hayim says: }{\marrow\sf #1}}
\def\micha#1{{\sc Micha says: }{\marrow\sf #1}}

\begin{titlepage}

\title{Semi-algebraic Range Reporting and Emptiness Searching 
with Applications\thanks{%
This work was supported by Grant 155/05 from the Israel Science Fund, 
by NSF Grants CCF-05-14079 and CCF-08-30272, by a grant from
the U.S.-Israel Binational Science Foundation, and by the Hermann
Minkowski--MINERVA Center for Geometry at Tel Aviv University.
This work is part of the second author's Ph.D. dissertation, prepared 
under the supervision of the first author at Tel Aviv University.} }

\author{
Micha Sharir\thanks{School of Computer Science, Tel Aviv University,
Tel Aviv 69978, Israel and Courant Institute of Mathematical
Sciences, New York University, New York, NY 10012, USA. {\sl
michas@post.tau.ac.il} }
\and
Hayim Shaul\thanks{School of Computer Science, Tel Aviv University,
Tel Aviv 69978, Israel. {\sl hayim@post.tau.ac.il} }
}

\maketitle

\begin{abstract}
\small
In a typical range emptiness searching (resp., reporting) problem, 
we are given a set $P$ of $n$ points in $\reals^d$, and wish to 
preprocess it into a data structure that supports efficient range 
emptiness (resp., reporting) queries, in which we specify a range 
$\sigma$, which, in general, is a semi-algebraic set in $\reals^d$ 
of constant description complexity, and wish to determine whether
$P\cap\sigma=\emptyset$, or to report all the points in $P\cap\sigma$.
Range emptiness searching and reporting arise in many
applications, and have been treated by Matou\v{s}ek~\cite{Ma:rph}
in the special case where the ranges are halfspaces bounded by
hyperplanes.
As shown in \cite{Ma:rph}, the two problems are closely related, and have
solutions (for the case of halfspaces) with similar performance bounds.
In this paper we extend the analysis to general
semi-algebraic ranges, and show how to adapt Matou\v{s}ek's technique,
without the need to {\em linearize} the ranges into
a higher-dimensional space. This yields more efficient solutions
to several useful problems, and we demonstrate the new technique
in four applications, with the following results:

\small
(i) An algorithm for ray shooting amid balls in $\reals^3$, which
uses $O(n)$ storage and $O^*(n)$ preprocessing,\footnote{%
  We use the notation $O^*(n^\gamma)$ to mean an upper bound of the
  form $C(\eps)n^{\gamma+\eps}$, which holds for any $\eps>0$,
  where $C(\eps)$ is a constant that depends on $\eps$.}
and answers a query in $O^*(n^{2/3})$ time, improving the previous
bound of $O^*(n^{3/4})$.

\small
(ii) An algorithm that preprocesses, in $O^*(n)$ time, a set $P$ of
$n$ points in $\reals^3$ into a data structure with $O(n)$ storage,
so that, for any query line $\ell$ (or, for that matter, any
simply-shaped convex set), the point of $P$
farthest from $\ell$ can be computed in $O^*(n^{1/2})$ time. This
in turn yields an algorithm that computes the largest-area triangle 
spanned by $P$ in time $O^*(n^{26/11})$, as well as nontrivial
algorithms for computing the largest-perimeter or largest-height
triangle spanned by $P$.

\small
(iii) An algorithm that preprocesses, in $O^*(n)$ time, a set $P$ of
$n$ points in $\reals^2$ into a data structure with $O(n)$ storage,
so that, for any query $\alpha$-fat triangle $\Delta$, we can
determine, in $O^*(1)$ time, whether $\Delta\cap P$ is empty.
Alternatively, we can report in $O^*(1) + O(k)$ time, the points
of $\Delta\cap P$, where $k=|\Delta\cap P|$.

\small
(iv) An algorithm that preprocesses, in $O^*(n)$ time, a set $P$ of
$n$ points in $\reals^2$ into a data structure with $O(n)$ storage,
so that, given any query semidisk $c$, or a circular cap larger than 
a semidisk, we can determine, in $O^*(1)$ time, whether $c\cap P$ 
is empty, or report the $k$ points in $c\cap P$ in
$O^*(1)+O(k)$ time.

\small
Adapting the recent techniques of \cite{AHP,AHPS,ArS}, we can turn
our solutions into 
efficient algorithms for approximate range counting (with small
relative error) for the cases mentioned above.

\small
Our technique is closely related to the notions of nearest- or
farthest-neighbor generalized Voronoi diagrams, and of the union
or intersection of geometric objects, where sharper bounds on the
combinatorial complexity of these structures yield faster range
emptiness searching or reporting algorithms. 
\end{abstract}

\end{titlepage}

\section{Introduction}
The main technical contribution of this paper is an extension of
Matou\v{s}ek's range emptiness and reporting data structures~\cite{Ma:rph}
(see also \cite{AM:dhr} for a dynamic version of the problem)
to the case of general semi-algebraic ranges. 

\paragraph{Ray shooting amid balls.}
A motivating application of
this study is ray shooting amid balls in $\reals^3$, where we want
to construct a data structure of linear size with near-linear
preprocessing,
which supports ray shooting queries in sublinear time. 
Typically, in problems of this sort, the bound on the
query time is some fractional power of $n$, the number of objects, and
the goal is to make the exponent as small as possible. For example,
ray shooting amid a collection of $n$ arbitrary triangles can be
performed in $O^*(n^{3/4})$ time (with linear storage)
\cite{AgM94}. Better solutions are known for various special
cases. For example, the authors have shown \cite{SS} that
the query time can be improved to $O^*(n^{2/3})$, when the triangles
are all {\em fat}, or are all stabbed by a common line.

At the other end of the spectrum, one is interested in ray shooting
algorithms and data structures where a ray shooting query can be
performed in logarithmic or polylogarithmic time (or even $O(n^\eps)$
time, for any $\eps>0$; this is $O^*(1)$ in our shorthand notation). 
In this case, the goal is to reduce the
storage (and preprocessing) requirements as much as possible.
For example, for arbitrary triangles (and even for the special case
of fat triangles), the best known bound for the storage requirement
(with logarithmic query time) is $O^*(n^{4})$ \cite{Ag91,AgM94}.
For balls, Mohaban and Sharir \cite{MS}, gave an algorithm with $O^*(n^{3})$
storage and $O^*(1)$ query time. However, when only linear
storage is used, the previously best known query time (for balls)
is $O^*(n^{3/4})$ (as in the case of general triangles).
In this paper we show, as an application of our general range
emptiness machinery, that this can be improved to $O^*(n^{2/3})$ time.

When answering a ray-shooting query for a set $S$ of input objects,
one generally reduces the problem to that of answering {\em segment
emptiness} queries, following the parametric searching scheme
proposed by Agarwal and Matou\v{s}ek~\cite{AgM93}
(see also Megiddo \cite{Meg} for the original underlying technique).

A standard way of performing the latter kind of queries is to
switch to a dual parametric space, where each object in the
input set is represented by a {\em point}. A segment $e$ in 
$\reals^3$ is mapped to a surface $\sigma_e$, which is the 
locus of all the points representing the objects that $e$ touches
(without penetrating into their interior).
Usually, $\sigma_e$ partitions the dual space into two portions,
one, $\sigma_e^+$, consisting of points representing objects
whose interior is intersected by $e$, and the other, 
$\sigma_e^-$, consisting of points representing objects 
that $e$ avoids. The segment-emptiness problem thus transforms
into a range-emptiness query: Does $\sigma_e^+$ contain any
point representing an input object?

\paragraph{Range reporting and emptiness searching.}
Range-emptiness queries of this kind have been studied by
Matou\v{s}ek~\cite{Ma:rph} (see also 
Agarwal and Matou\v{s}ek~\cite{AM:dhr}),
but only for the case where the ranges are
halfspaces bounded by hyperplanes. For this case, Matou\v{s}ek
has established a so-called {\em shallow-cutting lemma}, that shows the
existence of a $(1/s)$-cutting\footnote{%
  This is a partition of space (or a portion thereof) into a small
  number of simply-shaped cells, each of which is crossed by at most
  $n/s$ of the $n$ given surfaces (hyperplanes in this case).
  See below for more details.}
that covers the complement of the union of any $m$ given halfspace
ranges, whose size is significantly smaller than the size of a
$(1/s)$-cutting that covers the entire space. This lemma provides
the basic tool for partitioning a point set $P$, in the style of
\cite{Ma:ept}, so that {\em shallow} hyperplanes (those containing
at most $n/r$ points of $P$ below them, say, for some given 
parameter $r$) cross only a small number
of cells of the partition (see below for more details).
This in turn yields a data structure, known as a 
{\em shallow partition tree},
that stores a recursive partitioning of $P$, which enables us to
answer more efficiently halfspace range {\em reporting}
queries for shallow hyperplanes, 
and thus also halfspace range emptiness queries.
Using this approach, the query time (for emptiness)
improves from the general
halfspace range searching query cost of
$O^*(n^{1-1/d})$ to $O^*(n^{1-1/\lfloor d/2\rfloor})$.
Reporting takes $O^*(n^{1-1/\lfloor d/2 \rfloor} + k)$,
where $k$ is the output size.

Consequently, one way of applying this machinery for more 
general semi-algebraic ranges is to ``lift'' the 
set of points and the ranges into a higher-dimensional space by 
means of an appropriate {\em linearization}, as in \cite{AgM94},
and then apply the above machinery. 
(For this, one needs to assume that the given ranges have
{\em constant description complexity}, meaning that each range is a
Boolean combination of a constant number of polynomial equalities and
inequalities of constant maximum degree.
However, if the space in which the ranges are 
linearized has high dimension, the resulting range reporting or emptiness 
queries become significantly less efficient. Moreover, in many 
applications, the ranges are Boolean combinations of polynomial 
(equalities and) inequalities, which creates additional 
difficulties in linearizing the ranges, resulting in even worse running time. 

An alternative technique is to give up linearization, and instead work
in the original space.  As follows from the machinery of \cite{Ma:rph} 
(and further elaborated later in this paper), this requires, 
as a major tool, the (existence and) construction of a 
decomposition of the {\em complement of the union}
of $m$ given ranges (in the case of segment emptiness, these 
are the ranges $\sigma_e^+$, for an appropriate collection of 
segments $e$), into a small number of ``elementary cells'' 
(in the terminology of \cite{AgM94}---see also below).
Here we face, especially in higher dimensions,
a scarcity of sharp bounds on the complexity of the union
itself, to begin with, and then on the complexity of a decomposition
of its complement. Often, the best one can do is to decompose the
entire arrangement of the given ranges, which results in too many
elementary cells, and consequently in an algorithm
with poor performance. 

To recap, in the key technical step in answering general semi-algebraic 
range reporting or emptiness queries,
the best current approaches are either to construct a cutting of the
{\em entire} arrangement of the range-bounding surfaces in the
original space, or to construct a shallow cutting in another
higher-dimensional space into which the ranges can be linearized.
For many natural problems (including the segment-emptiness problem),
both approaches yield relatively poor performance.

As we will shortly note, in handling general semi-algebraic ranges,
we face another major technical issue, having to
do with the construction of efficient {\em test sets} of 
ranges (in the terminology of \cite{AgM94}, elaborated below). 
Addressing this issue is a major component of the analysis in this
paper, and is discussed in detail later on.

\paragraph{Our results.}
We propose a variant of the shallow-cutting machinery 
of~\cite{Ma:rph} for the case of semi-algebraic ranges, which avoids
the need for linearization, and works in the original space (which, 
for the case of ray shooting amid balls, is a 4-dimensional parametric
space in which the balls are represented as points).
While the machinery used by our variant is similar in principle to
that in \cite{Ma:rph}, there are several significant technical 
difficulties which require more careful treatment. 

Matou\v{s}ek's technique \cite{Ma:rph}, as well as ours, considers 
a finite set $Q$ of shallow ranges (called a {\em test set}), 
and builds a data structure which caters only for ranges 
in $Q$. Matou\v{s}ek shows how to build, for any given parameter $r$, 
a set of halfspaces of size polynomial in $r$, which represents well
{\em all} $(n/r)$-shallow ranges, in the following sense:
For any {\em simplicial partition}\footnote{%
  Briefly, this is a partition of $P$ into $O(r)$ subsets of roughly
  equal size, each enclosed by some simplex (in the linear case) 
  or some elementary cell (in the general semi-algebraic case);
  see~\cite{AgM94} and Section~\ref{section:range_emptiness} below.} 
$\Pi$ with parameter $r$, let $\kappa$ denote the maximal number 
of cells of $\Pi$ crossed by a halfspace in $Q$.
Then each $(n/r)$-shallow halfspace crosses at most $c\kappa$ cells of
$\Pi$, where $c$ is a constant that depends on the dimension.
Unfortunately (for the present analysis), 
the linear nature of the ranges is crucially needed for
the proof, which therefore fails for non-linear ranges.

Being a good representative of all shallow ranges, in the above sense,
is only one of the requirements from a good test set $Q$. The other
requirements are that $Q$ be small, so that, in particular,
it can be constructed efficiently, and that the (decomposition 
of the) complement of the union of any subset of $Q$ have small 
complexity. All these properties hold for the case of halfspaces 
bounded by hyperplanes, studied in \cite{Ma:rph}.

\label{testset_for_empty_only}
As it turns out, and hinted above,
obtaining a ``good'' test set $Q$ for general
semi-algebraic ranges, with the above properties, is not an easy task.
We give a simple general recipe for constructing such a set $Q$, but
it consists of more complex ranges than those in the original setup. 
A major problem with this recipe is that 
since the members of $Q$ have a more complex shape, 
it becomes harder to establish good bounds on the complexity of 
(the decomposition of) the
complement of the union of any subset of these generalized ranges.

Nevertheless, once a good test set has been shown to exist, and to 
be efficiently computable, it leads to a construction of an efficient
elementary-cell partition with a small crossing number for any empty
or shallow
original range. Using this construction recursively, one obtains a
{\em partition tree}, of linear size, so that any shallow original range
$\gamma$ visits only a small number of its nodes (where $\gamma$
visits a node if it {\em crosses} the elementary cell enclosing 
the subset of that node, meaning that it intersects this cell 
but does not fully contain it),
which in turn leads to an efficient range reporting or
emptiness-testing
procedure. This part, of constructing and searching the tree, is
almost identical to its counterparts in the earlier works
\cite{AgM94,Ma:rph,Ma:ept}, and we will not elaborate on it here, 
focusing only on the technicalities in the construction of a 
single ``shallow'' elementary-cell partition.

Developing all this machinery, and then putting it into action, we
obtain efficient data 
structures for the following applications, improving previous results or
obtaining the first nontrivial solutions. These instances are:

\paragraph{Ray shooting amid balls in 3-space.}
Given a set $S$ of $n$ balls in $\reals^3$, we construct, in 
$O^*(n)$ time, a data structure of $O(n)$ size, which can determine,
for a given query segment $e$, whether $e$ is empty (avoids all
balls), in $O^*(n^{2/3})$ time. Plugging this data structure into the
parametric searching technique of Agarwal and Matou\v{s}ek~\cite{AgM93},
we obtain a data structure for answering ray shooting queries amid the
balls of $S$, which has similar performance bounds.

We represent balls in 3-space as points in $\reals^4$, where a ball
with center $(a,b,c)$ and radius $r$ is mapped to the point
$(a,b,c,r)$, and each object $K\subset\reals^3$ is mapped to 
the surface $\sigma_K$, which is the locus of all (points 
representing) balls tangent to $K$ (i.e., balls that touch $K$,
but do not penetrate into its interior). In this case, 
the range of an object $K$ is the upper halfspace $\sigma_K^+$
consisting of all points lying above $\sigma_K$ (representing balls
that intersect $K$). The complement of the union of a subfamily of
these ranges is the region below the lower envelope of the
corresponding surfaces\footnote{%
  In our solution, we will use a test set of objects $K$ which are
  considerably more complex than just lines or segments, but are
  nevertheless still of constant description complexity.}
$\sigma_K$. The {\em minimization diagram} of this envelope is 
the 3-dimensional Euclidean Voronoi diagram of the corresponding 
set of objects.
Thus we reveal (what we regard as) a somewhat surprising connection 
between the problem of ray shooting amid balls and the problem of 
analyzing the complexity of Euclidean Voronoi diagrams of
(simply-shaped) objects in 3-space.

\paragraph{Farthest point from a line (or from any convex set) 
in $\reals^3$.}
Let $P$ be a set of $n$ points in $\reals^3$. We wish to preprocess
$P$ into a data structure of size $O(n)$, so that, for any query
line $\ell$, we can efficiently find the point of $P$ farthest from
$\ell$.  This is a useful routine for approximating polygonal paths 
in three dimensions; see \cite{daescu}.

As in the ray shooting problem, we can reduce such a query to a range
emptiness query of the form: Given a cylinder $C$, does it contain all
the points of $P$? (That is, is the complement of the cylinder empty?) 
We prefer to regard this as an instance 
of the complementary {\em range fullness} problem, which seeks 
to determine whether a query range is {\em full} (i.e., contains 
all the input points).

Our machinery can handle this problem. In fact, we can solve the 
range fullness problem for any family of {\em convex} ranges in
3-space, of constant description complexity.  Our solution 
requires $O(n)$ storage and near linear preprocessing, and 
answers a range fullness query in $O^*(n^{1/2})$ time,
improving the query time $O^*(n^{2/3})$ given by 
Agarwal and Matou\v{s}ek \cite{AgM94}. 

We then apply this result to solve the problem of finding the
largest-area triangle spanned by a set of $n$ points in 3-space.
The resulting algorithm requires $O^*(n^{26/11})$ time,
which improves a previous bound of $O^*(n^{13/5})$ due to Daescu and
Serfling \cite{daescu}.
We also adapt our machinery to compute efficiently the
largest-perimeter triangle and the largest-height triangle spanned by
such a point set.

In both this, and the preceding ray-shooting applications, we use the
general, more abstract recipe for constructing good test sets.

\paragraph{Fat triangle and circular cap range emptiness searching
and reporting.}
Finally, we consider two planar instances of the range emptiness 
and reporting problems, in which we are given a planar set $P$ of 
$n$ points, and the ranges are either {\em $\alpha$-fat triangles}
or sufficiently large {\em circular caps} (say, larger than a
semidisk). The general technique of Agarwal and 
Matou\v{s}ek~\cite{AgM94} yields, for any class of planar 
ranges with constant description complexity, a data structure 
with near linear preprocessing and linear storage, 
which answers such queries in time $O^*(n^{1/2})$ (for emptiness) 
or $O^*(n^{1/2}) + O(k)$ (for reporting).
We improve the query time to $O^*(1)$ and $O^*(1) + O(k)$,
respectively, in both cases.

In these planar applications, we abandon the general recipe, and
construct good test sets in an ad-hoc (and simpler) manner.
For $\alpha$-fat triangles (i.e., triangles with the property that
each of their angles is at least $\alpha$, which is some fixed 
positive constant), the test set consists of ``canonical''
$(\alpha/2)$-fat triangles, and the fast query performance is a
consequence of the fact that the complement of the union of
$m$ $\alpha'$-fat triangles is $O(m\log\log m)$, for any constant
$\alpha'>0$ \cite{MPSSW}. It is quite likely that our machinery can
also be applied to other classes of fat objects in the plane, for
which near-linear bounds on the complexity of their union are known
\cite{dB08,Ef05,EK99,ES00}. However, constructing a good test set for
each of these classes is not an obvious step.
We leave these extensions as open problems for further research.

For circular caps, the motivation for range emptiness searching
comes from the problem of finding, for a query consisting of a 
point $q$ and a line $\ell$, the point of $P$ which lies above 
$\ell$ and is nearest to $q$
(we only consider the case where $q$ lies on or above $\ell$).
Such a procedure was considered in \cite{DMSW}.
Using parametric searching, the latter problem can be reduced to that
of testing for emptiness of a circular cap centered at $q$ and
bounded by $\ell$ (the assumption on the location of $q$ ensures that
this cap is at least a semidisk). Here too we manage to construct a
test set which consists of (possibly slightly smaller) circular caps,
and we exploit the fact that the complexity of the union of $m$ such
caps is $O^*(m)$, as long as the caps are not too small (relative to
their bounding circles), to obtain the fast performance stated above.

\paragraph{Approximate range counting.}
Adapting the recent techniques of \cite{AHP,AHPS,ArS}, we can turn
our solutions into 
efficient algorithms for approximate range counting (with small
relative error) for the cases mentioned above.
That is, for a specified $\eps>0$, we can preprocess the input point
set $P$ into a data structure which can efficiently compute, for any
query range $\gamma$, an approximate count $t_\gamma$, satisfying
$(1-\eps)|P\cap\gamma| \le t_\gamma \le (1+\eps)|P\cap\gamma|$. The
performance of the resulting algorithms is detailed in
Section~\ref{sec:apx}. As observed in the papers just cited,
approximate range counting is closely related to the range emptiness
problem, which in fact is a special case of the former problem.
The algorithm in \cite{AHP} performs approximate range counting by a
randomized binary search over $|P\cap\gamma|$, where the search is 
guided by repeated calls to an emptiness testing routine on various
random samples of $P$. This algorithm uses emptiness searching as a
black box, so, plugging our solutions for this latter problem into
their algorithm, we obtain efficient approximate range counting
algorithms for the ranges considered in this paper. See
Section~\ref{sec:apx} for details.

\paragraph{Related work.}
Our study was originally motivated by work by Daescu and
others~\cite{DMSW,daescu} on path approximations and related problems.
In these applications one needs to compute efficiently the vertex of a
subpath which is farthest from a given segment (connecting the two
endpoints of the subpath). These works used the standard range searching
machinery of \cite{AgM94}, and motivated us to look for faster
implementations.

The general range emptiness (or reporting) problem was studied by the authors a
few years ago~\cite{ShSh2}. In this earlier version, we did not manage
to handle properly the issue of constructing a good test set, so the
results presented there are somewhat incomplete. The present paper
builds upon the previous one, but
provides a thorough analysis of this aspect of the problem, and
consequently obtains a complete and efficient solution to the problems
listed above, and lays down the foundation for obtaining efficient
solutions to many other similar problems---we believe indeed that the
applications given here only scratch the surface of the wealth of
potential future applications of this sort.

\section{Preliminaries and notations}
We begin with a brief review of the main concepts and notations
used in our analysis.

\paragraph{Range spaces.}
A range space is a pair $(X,\Gamma)$, where $X$ is a set and
$\Gamma\subseteq 2^X$ is a collection of subsets of $X$, called
{\em ranges}. In our
applications, $X=\reals^d$, and $\Gamma$ is a collection of
semi-algebraic sets of some specific type, each having
{\em constant description complexity}.
That is, each set in $\Gamma$ is given as a Boolean combination
of a constant number of polynomial equalities and inequalities
of constant maximum degree. To simplify the analysis, 
we assume\footnote{%
  This assumption is not essential, and is only made to simplify the
  presentation.},
as in \cite{AgM94}, that all the ranges in $\Gamma$ are defined by a
single Boolean combination, so that each polynomial $p$ in this
combination is $(d+t)$-variate, and each range $\gamma$ has $t$
degrees of freedom, so that if we substitute the values of these $t$
parameters into the last $t$ variables of each $p$, the resulting
Boolean combination defines the range $\gamma$.
This allows us to represent the ranges of $\Gamma$ as points in an
appropriate $t$-dimensional parametric space.

Under these special assumptions, the range space $(X,\Gamma)$ has
{\em finite VC-dimension}, a property formally defined in \cite{HW}.
Informally, it ensures that, for any finite subset $P$ of $X$, the
number of distinct ranges of $P$ is $O(|P|^\delta)$, where $\delta$ is
the VC-dimension.

As a matter of fact, we will consider
range spaces of the form $(P,\Gamma_P)$, where $P\subset\reals^d$
is a finite point set, and each range in $\Gamma_P$ is the
intersection of $P$ with a range in $\Gamma$.

\paragraph{Cuttings.}
Given a finite collection $\Gamma$ of $n$ semi-algebraic ranges in $\reals^d$,
as above, and a parameter $r<n$, a {\em $(1/r)$-cutting} for $\Gamma$
is a partition $\Xi$ of $\reals^d$ (or of some portion of $\reals^d$)
into a finite number of relatively open cells of dimensions
$0,1,\ldots,d$,
so that each cell is {\em crossed} by at most $n/r$ ranges of $\Gamma$,
where a range $\gamma\in\Gamma$ is said to cross a cell $\sigma$ if
$\gamma\cap\sigma\ne\emptyset$, but $\gamma$ does not fully contain
$\sigma$. We will also need to consider {\em weighted} $(1/r)$-cuttings,
where each range $\gamma\in\Gamma$ has a positive weight $w(\gamma)$,
and each cell of $\Xi$ is crossed by ranges whose total weight is at most
$W/r$, where $W=\sum_{\gamma\in\Gamma} w(\gamma)$ is the overall weight
of all the ranges in $\Gamma$.

\paragraph{Shallow ranges.}
A range $\gamma\in\Gamma$ is called {\em $k$-shallow} with respect to
a set $P$ of points in $\reals^d$, if $|\gamma\cap P|\le k$.

\paragraph{Elementary cells.}
Define, as in \cite{AgM94}, an {\em elementary cell} in $\reals^d$
to be a connected relatively open semi-algebraic set of some dimension
$k\le d$, which is homeomorphic to a ball
and has constant description complexity.
As above, we assume, for simplicity,
that the elementary cells are defined by a single
Boolean combination involving $t$ free variables, and each cell is
determined by fixing the values of these $t$ parameters.

\paragraph{Elementary cell partition.}
Let $P$ be a set of $n$ points in $\reals^d$.
An {\em elementary cell partition} of $P$ is a collection
$\Pi = \{(P_1, s_1), \ldots, (P_m, s_m)\}$, for some integer $m$,
such that (i) $\{P_1,\ldots,P_m\}$ is a partition of $P$ (into pairwise
disjoint subsets), and (ii) each $s_i$ is an elementary cell that contains
the respective subset $P_i$. In general, the cells $s_i$ need not be disjoint.
Usually, one also specifies a parameter $r\le n$, and requires that
$n/r \le |P_i| \le 2n/r$ for each $i$, so $m=O(r)$.

\paragraph{The function $\zeta(r)$.}
In Lemma~\ref{lemma:shallow_cutting} and Theorem~\ref{theo:part},
we use a function $\zeta(r)$ that bounds the number of elementary 
cells in a decomposition of the complement of the union of any
$r$ ranges of $\Gamma$. We assume
that $\zeta(r)$ is ``well behaved'', in the sense that for each $c>0$
there exists $c'>0$ such that $\zeta(cr)\le c'\zeta(r)$ for every $r$.
We also assume that $\zeta(r) = \Omega(r)$.

\paragraph{$(\nu,\alpha)$-samples and shallow $\eps$-nets.}
We recall the result of Li et al.~\cite{LLS01}, and adapt it, similar
to the recent observations in \cite{HPS}, to obtain a useful extension
of the notion of $\eps$-nets.

Let $(X,\R)$ be a range space of finite VC-dimension $\delta$,
and let $0<\alpha,\nu <1$ be two given parameters.
Consider the distance function
$$
d_\nu(r,s) = \frac{|r-s|}{r+s+\nu}, \quad\quad \mbox{for $r,s\ge 0$} .
$$
A subset $N\subseteq X$ is called a {\em $(\nu,\alpha)$-sample} if
for each $R\in\R$ we have
$$
d_\nu\left( \frac{|X\cap R|}{|X|}, \frac{|N\cap R|}{|N|} \right) < \alpha .
$$
\begin{theorem}[Li et al.~\cite{LLS01}] \label{lls}
A random sample $N$ of 
$$
O\left( \frac{1}{\alpha^2\nu} \left( \delta \log\frac{1}{\nu} +
\log\frac{1}{q} \right) \right)
$$
elements of $X$ is a $(\nu,\alpha)$-sample with probability 
at least $1-q$.
\end{theorem}

Har-Peled and Sharir~\cite{HPS}
show that, by appropriately choosing $\alpha$ and $\nu$, various
standard constructs, such as $\eps$-nets and $\eps$-approximations,
are special cases of $(\nu,\alpha)$-samples. Here we follow a 
similar approach, and show the existence of small-size 
{\em shallow $\eps$-nets}, a new notation introduced in this paper.

Let us first define this notion. 
Let $(X,\R)$ be a range space of finite VC-dimension $\delta$,
and let $0<\eps<1$ be a given parameter.  A subset $N\subseteq X$ 
is a shallow $\eps$-net if it satisfies the following two properties,
for some absolute constant $c$.

\noindent
(i) For each $R\in \R$ and for any parameter $t\ge 0$, 
if $|N\cap R| \le t\log\frac{1}{\eps}$ then
$|X\cap R| \le c(t+1)\eps|X|$.

\noindent
(ii) For each $R\in \R$ and for any parameter $t\ge 0$, 
if $|X\cap R| \le t\eps|X|$ then
$|N\cap R| \le c(t+1) \log\frac{1}{\eps}$.

Note the difference between shallow and standard $\eps$-nets:
Property (i) (with $t=0$) implies that a shallow $\eps$-net is
also a standard $\eps$-net (possibly with a recalibration of
$\eps$). Property (ii) has no parallel in the case of standard
$\eps$-nets -- there is no guarantee how a standard net interacts
with small ranges.

\begin{theorem} \label{epsshallow}
A random sample $N$ of 
$$
O\left( \frac{1}{\eps} \left( \delta \log\frac{1}{\eps} +
\log\frac{1}{q} \right) \right)
$$
elements of $X$ is a shallow $\eps$-net with probability at least 
$1-q$. 
\end{theorem}

\noindent{\bf Proof:}
Take $\alpha=1/2$, say, and calibrate the constants in the size of $N$
to guarantee, with probability $1-q$, that $N$ is an
$(\eps,1/2)$-sample. Assume that this is indeed the case.
For a range $R\in\R$, put $X_R=|X\cap R|/|X|$ and $N_R=|N\cap R|/|N|$.
We have
$$
d_\eps(X_R,N_R) = \frac{|X_R-N_R|}{X_R+N_R+\eps} < \frac12 .
$$
That is,
$$
|X_R-N_R| < \frac12(X_R+N_R+\eps) ,
$$
or
$$
X_R < 3N_R + \eps , \quad \mbox{and, symmetrically,} \quad
N_R < 3X_R + \eps .
$$
This is easily seen to imply properties (i) and (ii). For (i),
let $R$ be a range for which $|N\cap R| \le t\log\frac{1}{\eps}$; 
that is, $N_R \le \beta t\eps$, for some absolute constant $\beta$
(proportional to the VC-dimension). Then
$$
|X\cap R| = |X|\cdot X_R < |X|(3N_R + \eps) \le (3\beta t+1)\eps |X| .
$$
For (ii), let $R$ be a range for which $|X\cap R| \le t\eps|X|$;
that is, $X_R \le t\eps$. Then
$$
|N\cap R| = |N|\cdot N_R < |N|(3X_R+\eps) \le (3t+1)\eps|N| \le
(3t+1)\gamma \log\frac{1}{\eps} ,
$$ 
for another absolute constant $\gamma$ (again, proportional to the
VC-dimension).
$\Box$

\section{Semi-algebraic range reporting or emptiness searching}
\label{section:range_emptiness}
 
\paragraph{Shallow cutting in the semi-algebraic case.}
We begin by extending the shallow cutting lemma of 
Matou\v{s}ek~\cite{Ma:rph} to the more general setting of 
semi-algebraic ranges. This extension is fairly straightforward, 
although it involves several technical steps that deserve to be
highlighted.

\label{app:not}

\begin{lemma}[Extended Shallow Cutting Lemma]
\label{lemma:shallow_cutting}
Let $\Gamma$ be a collection of $n$ semi-algebraic ranges in $\reals^d$.
Assume that the complement of the union of any subset of
$m$ ranges in $\Gamma$ can be decomposed into at most 
$\zeta(m)$ elementary cells, for a well-behaved function $\zeta$ as
above.
Then, for any $r\le n$, there exists a ($1/r$)-cutting $\Xi$
with the following properties:\\
{(i)} The union of the cells of $\Xi$ contains the complement of the union
of $\Gamma$.\\
{(ii)} $\Xi$ consists of $O(\zeta(r))$ elementary cells.\\
{(iii)} The complement of the union of the cells of $\Xi$ is
contained in a union of $O(r)$ ranges in $\Gamma$.
\end{lemma}

\begin{figure}[htbp]
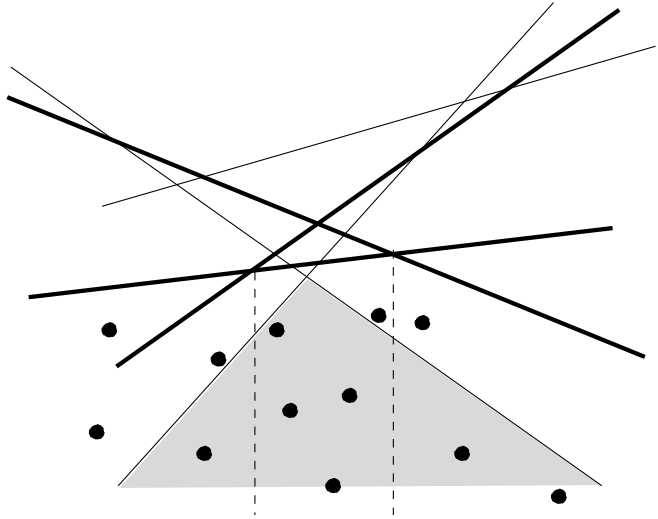

\begin{center}
\input shallow_lemma.pstex_t
\caption{A planar point set and a collection $\Gamma$ of upper
halfplanes. A random sample of the lines bounding these ranes is shown
in bold, with a decomposition of the region below their lower
envelope, which contains the region below the lower envelope of all
the bounding lines, drawn shaded.}
\label{fig:shallow-lemma}
\end{center}
\end{figure}

See Figure \ref{fig:shallow-lemma} for an illustration.

\noindent{\it Proof.}
The proof is a fairly routine adaptation of the proof in \cite{Ma:rph}.
We employ a variant of the method of Chazelle and Friedman \cite{CF} for
constructing the cutting. Let $\Gamma'$ be a random sample of $O(r)$ ranges
of $\Gamma$, and let $E'$ denote the complement of the union of $\Gamma'$. By
assumption, $E'$ can be decomposed into at most $O(\zeta(r))$ elementary
cells. The resulting collection $\Xi$ of these cells is such that
their union clearly contains the complement of the union of $\Gamma$.
Moreover, the complement of the union of $\Xi$ is the union of the
$O(r)$ ranges of $\Gamma'$. Hence, $\Xi$ satisfies all three conditions
(i)--(iii), but it may fail to be a $(1/r)$-cutting. 

This latter property is enforced as in \cite{CF}, by further 
decomposing each cell $\tau$ of $\Xi$ that is crossed by more than 
$n/r$ ranges of $\Gamma$, using additional subsamples
from the surfaces that cross $\tau$.  
Specifically, for each cell $\tau$ of $\Xi$, let $\Gamma_\tau$ denote
the subset of those ranges in $\Gamma$ that cross $\tau$, and put
$\xi_\tau = |\Gamma_\tau|r/n$. If $\xi_\tau > 1$, we sample
$q=O(\xi_\tau\log\xi_\tau)$ ranges from $\Gamma_\tau$,
construct the complement of the union of these ranges, decompose it
into at most $\zeta(q)$ elementary cells, and clip them to within $\tau$.
The resulting collection $\Xi'$ of subcells, over all cells $\tau$ of
the original $\Xi$, clearly satisfies (i). The analysis of \cite{CF}
(see also \cite{AMS98})
establishes an exponential decay property on the number of cells of
$\Xi$ that are crossed by more than $\xi n/r$ ranges, as a function of
$\xi$. Specifically, as in \cite{AMS98}, the expected number of such 
cells is $O(2^{-\xi}\EE(\zeta(|\Gamma''|))$, where $\Gamma''$ is 
another random sample of $\Gamma$, where each member of $\Gamma$ 
is chosen with probability $\frac{r}{n\xi}$.
This property implies, as usual \cite{CF}, that $\Xi'$ is
(with high probability) a $(1/r)$-cutting, and it also implies that
the size of $\Xi'$ is still $O(\zeta(r))$, assuming $\zeta$ to be well
behaved. Since we have only refined
the original cells of $\Xi$, the number of ranges that cover the 
complement of the union of the final cells is still $O(r)$.
$\Box$

A special case that arises frequently is where each range in $\Gamma$
is an upper (or lower) halfspace bounded by the graph of some continuous
$(d-1)$-variate function. In this case the complement $K$ of the union
of $r$ ranges is the portion of space that lies below the
{\em lower envelope} of the bounding graphs. In this case, it suffices
to decompose the graph of the lower envelope itself into at most
$\zeta(r)$ elementary cells. Indeed, having done that, we can extend
each cell $\tau$ within the envelope into the cell $\tau^-$ consisting
of all points that lie vertically below $\tau$. The new cells decompose
$K$ and are also elementary.

As already discussed in the introduction, 
obtaining tight or nearly tight bounds for $\zeta(r)$ is still a major
open problem for many instances of the above setup. For example,
decomposing an upper envelope of $r$ $(d-1)$-variate functions of
constant description complexity into $O^*(r^{d-1})$ elementary cells
is still open for any $d\ge 4$. (This bound is best possible in the
worst case, since
it is the worst-case tight bound on the complexity of such an
undecomposed envelope~\cite{SA}.) The cases $d=2$ (upper envelope of
curves in the plane) and $d=3$ (upper envelope of 2-dimensional
surfaces in 3-space) are easy. In these cases $\zeta(r)$ is
proportional to the complexity of the envelope, which in the worst
case is near-linear for $d=2$ and near-quadratic for $d=3$ \cite{SA}.
In higher dimensions, the only general-purpose bound known to date is
the upper bound obtained by computing the {\em vertical decomposition} of
the {\em entire} arrangement of the given surfaces, and extracting
from it the relevant cells that lie on or above the envelope. In
particular, for $d=4$ the bound is $\zeta(r)=O^*(r^{4})$, as
follows from the results of \cite{Kol}. This leaves a gap of
about a factor of $r$ between this bound and the bound $O^*(r^{3})$
on the complexity of the undecomposed envelope. Of course, in certain
special cases, most notably the case of hyperplanes, as studied in
\cite{Ma:rph}, both the envelope and its decomposition have
(considerably) smaller complexity.

The situation with the complexity of the union of geometric objects
is even worse. While considerable progress was recently made on many 
special cases in two and three dimensions (see \cite{APS} for a recent
comprehensive survey), there are only very few sharp bounds on the
complexity of unions in higher dimensions. Worse still, even when a
sharp bound on the complexity of the union is known, obtaining
comparable bounds on the complexity of a decomposition of the
complement of the union is a much harder problem (in $d\ge 3$
dimensions). As an example, the union of $n$ congruent infinite
cylinders in 3-space is known to have near-quadratic
complexity~\cite{AS00}, but it is still an open problem whether its
complement can be decomposed into a near-quadratic number of
elementary cells.

\paragraph{Partition theorem for shallow semi-algebraic ranges.}
We next apply the new shallow cutting lemma to construct an
elementary cell partition of a given input point set $P$, with respect 
to a specific set $Q$ of ranges. This is done in a fairly similar 
way to that in~\cite{AgM94} (see also \cite{Ma:rph,Ma:ept}).
A major difference in handling the semi-algebraic case is the
construction of a set $Q$ of ranges that will be (a) small enough, 
and (b) representative of all shallow (or empty) ranges, in a sense
discussed in detail below. The method given in \cite{AgM94} does not
work in the general semi-algebraic case, and different, 
sometimes ad-hoc approaches need to be taken.

The following theorem summarizes the main part of the construction
(except for the construction of $Q$).

\begin{theorem}[Extended Partition Theorem]
\label{theo:part}
Let $P$ be a set of $n$ points in $\reals^d$, let $\Gamma$ be
a family of semi-algebraic ranges of constant description complexity,
and let $r$ be fixed. Let $Q$ be another finite collection 
(not necessarily a subset of $\Gamma$) of 
semi-algebraic ranges of constant description complexity with the
following properties: 
(i) The ranges in $Q$ are all $(n/r)$-shallow.
(ii) The complement of the union of any $m$ ranges of $Q$
can be decomposed into at most $\zeta(m)$ elementary cells, 
for any $m$. 
(iii) Any $(n/r)$-shallow range $\gamma \in \Gamma$ can be covered
by the union of at most $\delta$ ranges of $Q$, where $\delta$ is a
constant.  \\
Then there exists an elementary cell partition
$\Pi$ of $P$, of size $O(r)$, into subsets of size roughly $n/r$,
such that the crossing number of
any $(n/r)$-shallow range in $\Gamma$ with the cells of $\Pi$
is either $O(r/\zeta^{-1}(r)+\log r\log |Q|)$, if
$\zeta(r)=\Omega(r^{1+\eps})$, for any fixed $\eps>0$, or
$O(r\log r/\zeta^{-1}(r)+\log r\log |Q|)$, otherwise.
\end{theorem}

the proof, which, again, is similar to those in
\cite{AgM94,Ma:rph,Ma:ept}, 
proceeds through the following steps.
We first have:
\begin{lemma}
\label{lemma:half_partition}
Let $P$ be a set of $n$ points in $\reals^d$, and $r < n$ a parameter.
Let $Q$ be a set of $(n/r)$-shallow ranges, with the property that the
complement of the union of any subset of $m$ ranges of $Q$ can be
decomposed into at most $\zeta(m)$ elementary cells, for any $m$.
Then there exists a subset $P' \subseteq P$ of at least $n/2$
points and an elementary cell partition
$\Pi = \{(P_1, s_1), \ldots, (P_m, s_m)\}$ for $P'$ with
$|P_i| = \lfloor n/r \rfloor$ for all $i$, such that each
range of $Q$ crosses at most $O(r/\zeta^{-1}(r) + \log|Q|)$
cells $s_i$ of $\Pi$.
\end{lemma}

\noindent{\it Proof.}
We will inductively construct disjoint sets $P_1,\ldots,P_m \subset P$
of size $n/r$ and elementary cells $s_1, \ldots, s_m$
such that $P_i \subseteq s_i$ for each $i$.
The construction terminates when $|P_1 \cup \cdots \cup P_m| \ge n/2$.
Suppose that $P_1, \ldots, P_{i-1}$ have already been constructed, and
set $P'_i := P\setminus\bigcup_{j<i} P_j$. We construct $P_i$ 
as follows: For a range $\sigma \in Q$, let $\kappa_i(\sigma)$
denote the number of cells among $s_1, \ldots, s_{i-1}$ crossed
by $\sigma$. We define a weighted collection $(Q,w_i)$ of ranges, so
that each range $\sigma \in Q$ appears with weight (or multiplicity)
$w_i(\sigma) = 2^{\kappa_i(\sigma)}$. 
We put $w_i(Q)=\sum_{\sigma \in Q} w_i(\sigma)$.
By Lemma~\ref{lemma:shallow_cutting} and by our assumption that the
function $\zeta(r)$ is well behaved, there exists a $(1/t)$-cutting
$\Xi_i$ for the weighted collection $(Q,w_i)$ of size at most $r/4$,
for an appropriate choice of $t=\Theta(\zeta^{-1}(r))$, with the
following properties: The union of $\Xi_i$ contains the complement
of the union of $Q$, and the complement of the union of $\Xi_i$ is
contained in the union of $O(t)$ ranges of $Q$. Since all these ranges
are $(n/r)$-shallow, the number of points of $P$ not in the union of
$\Xi_i$ is at most $O(t)\cdot (n/r) = n \cdot O(\zeta^{-1}(r)/r)$,
and our assumptions 
on $\zeta(r)$ imply that this is smaller than $n/4$, if we choose 
$t$ appropriately.
Since we assume that $|P'_i|\ge n/2$, it follows that
at least $n/4$ points of $P'_i$ lie in the union of the at 
most $r/4$ cells of $\Xi_i$. By the pigeonhole principle,
there is a cell $s_i$ of $\Xi_i$ containing at least $n/r$
points of $P'_i$. We take $P_i$ to be some subset of
$P'_i \cap s_i$ of size exactly $n/r$, and make $s_i$ the cell
in the partition which contains $P_i$.

We next establish the asserted bound on the crossing numbers 
between the ranges of $Q$ and the elementary cells 
$s_1,\ldots,s_m$, in the following standard manner. 
The final weight $w_m(\sigma)$ of a range 
$\sigma \in Q$ with crossing number $\kappa$ (with respect to the
final partition) is $2^\kappa$. On the other hand, each
newly added cell $s_i$ is crossed by ranges of $Q$ of total weight
$O(w_i(Q)/\zeta^{-1}(r))$, because $s_i$ is an 
elementary cell of the corresponding weighted $(1/t)$-cutting 
$\Xi_i$. The weight of each of these crossing ranges is doubled 
at the $i$-th step, and the weight of all the other ranges 
remains unchanged. Thus
$w_{i+1}(Q) \le w_i(Q)(1 + O(1/\zeta^{-1}(r)))$.
Hence, for each range $\sigma\in Q$ we have
$$
w_m(\sigma) \le w_m(Q) \le
|Q|\left(1+O\left(\frac{1}{\zeta^{-1}(r)}\right)\right)^m \le
|Q|\left(1+O\left(\frac{1}{\zeta^{-1}(r)}\right)\right)^{O(r)} \le
|Q|e^{O(r/\zeta^{-1}(r))} ,
$$
and thus $\kappa = \log w_m(\sigma) = O(r/\zeta^{-1}(r) + \log |Q|)$.
$\Box$

\paragraph{Discussion.}
The limitation of Lemma~\ref{lemma:half_partition} is that the bound
that it derives (a) applies only to ranges in $Q$, and (b) includes
the term $\log |Q|$. An ingenious component of the analysis in
\cite{Ma:rph} overcomes both problems, by choosing a test set $Q$ of
ranges whose size is only polynomial in $r$ (and, in particular,
is independent of $n$), which is nevertheless sufficiently
representative of all shallow ranges, in the sense that the crossing
number of any $(n/r)$-shallow range is 
$O(\max\{\kappa(\sigma) \mid \sigma\in Q\})$. This implies that
Lemma~\ref{lemma:half_partition} holds for {\em all} shallow ranges,
with the stronger bound which does not involve $\log |Q|$.

Unfortunately, the technique of \cite{Ma:rph} does not extend to the
case of semi-algebraic ranges, as it crucially relies on the linearity
of the ranges.\footnote{%
  It uses point-hyperplane duality, and exploits the fact that a
  halfspace (bounded by a hyperplane) intersects a simplex if and only
  if it contains a vertex of the simplex, which is false in the
  general semi-algebraic case.}
The following lemma gives a sufficient condition for 
a test set $Q$ to be representative of the relevant shallow ranges, in the
sense that $Q$ satisfies the assumptions made
in Theorem~\ref{theo:part}.
That is:

\begin{lemma} \label{lemma:const_cover}
Let $P$ be a set of $n$ points in $\reals^d$, and let $\Gamma$ 
be a family of semi-algebraic ranges with constant description 
complexity. Consider an elementary-cell partition 
$\Pi = \{ (P_1,s_1),\ldots,(P_r,s_r) \}$ of $P$ such that 
$|P_i| = n/r$ for each $i$.
Let $Q$ be a finite set of $(n/r)$-shallow ranges (not 
necessarily ranges of $\Gamma$), so that the maximal crossing 
number of a range $q \in Q$ with respect to $\Pi$ is $\kappa$. 
Then, for any range $\gamma\in\Gamma$ which is contained in the 
union of at most $\delta$ ranges of $Q$ (for some constant
$\delta$), 
the crossing number of $\gamma$ is at most $(\kappa+1)\delta$.
\end{lemma}

\noindent{\it Proof.}
Let $\gamma\in\Gamma$ be a range for which there exist
$\delta$ ranges $q_1,\ldots,q_\delta$ of $Q$ such that 
$\gamma \subseteq q_1 \cup \cdots \cup q_\delta$. Then, if
$\gamma$ crosses a cell $s_i$ of $\Pi$, then at least one of the
covering ranges $q_j$ must either cross $s_i$ or fully contain $s_i$.
The number of cells of $\Pi$ that can be crossed by any single $q_j$
is at most $\kappa$, and each $q_j$ can fully contain at most one cell
of $\Pi$ (because $q_j$ is $(n/r)$-shallow).\footnote{%
  By choosing a slightly smaller value for $r$ in the construction
  of the partition, we can even rule out the possibility that a 
  range $q_j$ fully contains a cell of $\Pi$. This however has no 
  effect on the asymptotic bounds that the analysis derives.}
Hence, the overall number
of cells of $\Pi$ that $\gamma$ can cross is at most
$(\kappa+1)\delta$, as asserted.
$\Box$

\paragraph{Proof of Theorem~\ref{theo:part}.}
Apply Lemma~\ref{lemma:half_partition} to the input set
$P_0 = P$, with parameter $r_0 = r$. This yields an 
elementary-cell partition $\Pi_0$ for (at least) half of the 
points of $P_0$, which satisfies the properties of that lemma. 
Let $P_1$ denote the set of the remaining points of $P_0$, and 
set $r_1 = r_0/2$. Apply Lemma~\ref{lemma:half_partition} again 
to $P_1$ with parameter $r_1$, obtaining an elementary cell 
partition $\Pi_1$ for (at least) half of the points of $P_1$. 
We iterate this process $k = O(\log r)$ times,
until the set $P_k$ has fewer than $n/r$ points. We take $\Pi$ to
be the union of all the elementary-cell partitions $\Pi_i$ formed
so far, together with one large cell containing all the remaining
points of $P_k$. The resulting elementary-cell partition of $P$
consists of at most $1+r+r/2+r/4+\ldots \le 2r$ subsets, each 
of size at most $n/r$. The crossing number of a range in $Q$ is,
by Lemma~\ref{lemma:half_partition},
$$
O\left(
\sum_{i=1}^{\log r} \left(
(r/2^i)/\zeta^{-1}(r/2^i) + \log |Q| \right) \right) .
$$
Our assumptions on $\zeta$ imply that if
$\zeta(r)=\Omega(r^{1+\eps})$, for any fixed $\eps>0$, the first
terms add up to $O(r/\zeta^{-1}(r))$; otherwise we can bound their sum
by $O(r\log r/\zeta^{-1}(r))$.
Hence, by the properties of $Q$ and by Lemma~\ref{lemma:const_cover},
the crossing number of any empty range is also
$O(r/\zeta^{-1}(r) + \log |Q|\log r)$ or
$O(r\log r/\zeta^{-1}(r) + \log |Q|\log r)$, respectively.
$\Box$

\paragraph{Partition trees and reporting or emptiness searching.}
As in the classical works on range
searching~\cite{AgM94,Ma:rph,Ma:ept}, we apply Theorem~\ref{theo:part}
recursively, and obtain a {\em partition tree} $\T$, where each node
$v$ of $\T$ stores a subset $P_v$ of $P$ and an elementary cell $\sigma_v$
enclosing $P_v$. The children of a node $v$ are obtained from an
elementary cell partition of $P_v$---each of them stores one of the
resulting subsets of $P_v$ and its enclosing cell.
At the leaves, the size of the subset that is stored is $O(r)$.

Testing a range $\gamma$ for emptiness is done by searching with 
$\gamma$ in $\T$. At each visited node $v$, where
$\gamma\cap\sigma_v\ne\emptyset$, we test whether 
$\gamma\supseteq\sigma_v$, in which case $\gamma$ is not empty.
Otherwise, we find the children of $v$ whose cells are intersected by
$\gamma$. If there are too many of them we know that $\gamma$ is not
empty. Otherwise, we recurse at each child.

Reporting is performed in a similar manner. If $\sigma_v\subseteq\gamma$ we
output all of $\sigma_v$. Otherwise, we find the children of $v$ whose
cells are intersected by $\gamma$. If there are too many of them we know that
$\gamma$ is not $(n_v/r)$-shallow (with respect to $P_v$), so, if $r$ is a
constant, we can afford to check every element of $P_v$ for containment
in $\gamma$, and output those points that do lie in $\gamma$. If there are
not too many children, we recurse in each of them.

The efficiency of the search depends on the function $\zeta(m)$. If
$\zeta(m)=O^*(m^k)$ then an emptiness query takes $O^*(n^{1-1/k})$ time,
and a reporting query takes $O^*(n^{1-1/k}) +O(t)$, where
$t$ is the output size. Thus
making $\zeta$ (i.e., $k$) small is the main challenge in this
technique.

\paragraph{A general recipe for constructing good test sets.}
Let $\Gamma$ be the given collection of semi-algebraic ranges of
constant description complexity. As above, we assume that each range
$\gamma\in\Gamma$ has $t$ degrees of freedom, for some constant
parameter $t$, so it can be represented as a point $\gamma^*$ in a
$t$-dimensional parametric space, which, for convenience, we 
denote as $\reals^t$. Each input point $p\in P$ is mapped to a region
$K_p$, which is the locus of all points representing ranges which
contain $p$.

We fix a parameter $r\ge 1$, and choose a random sample $N$ of
$ar\log r$ points of $P$, where $a$ is a sufficiently large constant.
We form the set 
$N^* = \{K_p \mid p\in N\}$, construct the arrangement $\A(N^*)$, and
let $V=A_{\leq k}(N^*)$ denote the region consisting of all points contained
in at most $k$ ranges of $N^*$, where $k=b\log r$ and $b$ is an absolute
constant that we will fix later. We
decompose $V$ into elementary cells, using,
e.g., vertical decomposition \cite{SA}. In the worst case, we get
$O^*(r^{2t-4})$ elementary cells \cite{CEGS,Kol}.
\footnote{%
  Here, in this dual construction, we do not need any sharper bound;
  any bound polynomial in $r$ is sufficient for our purpose.}

Let $\tau$ be one of these cells. We associate with $\tau$ 
a generalized range $\gamma_\tau$ in $\reals^d$, which is the union
$\bigcup\{\gamma \mid \gamma^*\in\tau\}$. Since $\tau$ has constant
description complexity, as do the ranges of $\Gamma$, it is easy to
show that $\gamma_\tau$ is also a semi-algebraic set of constant
description complexity (see~\cite{BPR06}). 

We define the test set $Q$ to consist of all the generalized ranges 
$\gamma_\tau$, over all cells $\tau$ in the decomposition of
$V$, and claim that, with high probability (and with an appropriate choice
of $b$), $Q$ is a good test set, in the following three aspects.

\noindent
(i) {\bf Compactness.}
$|Q|=O^*(r^{2t-4})$; that is, the size of $Q$ is polynomial in $r$
and independent of $n$.

\noindent
(ii) {\bf Shallowness.}
Each range $\gamma_\tau$ in $Q$ is
$\beta(n/r)$-shallow with respect to $P$, for some constant
parameter $\beta$.

\noindent
(iii) {\bf Containment.}
Every {\em $(n/r)$-shallow} range $\gamma\in\Gamma$ is contained in a
single range $\gamma_\tau$ of $Q$.

Property (i) is obvious.
Consider the range space $(P, \Gamma^*)$, where $\Gamma^*$ consists of all
generalized ranges $\gamma_\tau$, over all elementary cells $\tau$ of the
form arising in the above vertical decomposition. It is a fairly easy
exercise to show that $(P, \Gamma^*)$ also has finite VC-dimension. See,
e.g., \cite{SA}. By Theorem~\ref{epsshallow}, if $a$ is a sufficiently
large constant (proportional to the VC-dimension of $(P, \Gamma^*)$) then
$N$ is a shallow $(1/r)$-net for both range spaces $(P, \Gamma)$ and
$(P, \Gamma^*)$, with high probability, so we assume that $N$ is indeed
such a shallow $(1/r)$-net.

Let $\gamma_\tau \in Q$. Note that any point $p \in P$ in 
$\gamma_\tau$ lies in a range $\gamma\in\Gamma$ with
$\gamma^*\in\tau$. By definition, $\gamma^*$ also belongs to $K_p$,
and so $K_p$ crosses or fully contains $\tau$. 
Since $\tau$ is $(b\log r)$-shallow in  $\A(N^*)$, it is 
fully contained in at most $b\log r$ regions $K_p$,
for $p \in N$ (and is not crossed by any such
region). Hence $|\gamma_\tau \cap N| < b\log r$, so, since $N$ is a shallow
$(1/r)$-net for $(P, \Gamma^*)$, we have $|\gamma_\tau \cap P| < c(b+1)n/r$,
so $\gamma_\tau$ is $(c(b+1)n/r)$-shallow, which establishes (ii).

For (iii), let $\gamma \in \Gamma$ be an $(n/r)$-shallow range. Since
$N$ is a shallow $(1/r)$-net for $(P, \Gamma)$, and
$|\gamma \cap P| \leq |P|/r$, we have $|\gamma \cap N| \leq 2c\log r$.
Hence, with $b \geq 2c$, $\gamma \in V$, so there is a cell $\tau$
of the decomposition which contains $\gamma$, which, by construction,
implies that $\gamma \subseteq \gamma_\tau$, thus establishing (iii).

To make $Q$ a really good test set, we also need the following
fourth property:

\noindent
(iv) {\bf Efficiency.}
There exists a good bound on the associated function 
$\zeta(m)$, bounding the size of a decomposition
of the complement of the union of any $m$ ranges of $Q$. 

The potentially rather complex shape of these generalized ranges makes 
it harder to obtain, in general, a good bound on $\zeta$. 

In what follows, we manage to use this general recipe in two of our four
applications (ray shooting amid balls and range fullness searching),
with good bounds on the corresponding functions $\zeta(\cdot)$. In two
other planar applications (range emptiness searching with fat triangles and
with circular caps), we abandon the general technique, and construct ad hoc
good test sets.

\paragraph{Remark:}
In the preceding construction, we wanted to make sure that every
$(n/r)$-shallow range $\gamma \in \Gamma$ is covered by a range of $Q$.
If we only need this property for {\em empty} ranges $\gamma$ (which is
the case for emptiness testing), it suffices to consider only the
$0$-level of $\A(N^*)$, i.e., the complement of the union of $N^*$.
Other than this simplification, the construction proceeds as above.

\section{Fullness searching and reporting outliers for convex ranges}

Let $P$ be a set of $n$ points in 3-space, and let $\Gamma$ 
be a set of convex ranges of constant description complexity. 
We wish to preprocess $P$ in near-linear time into a data structure
of linear size,
so that, given a query range $\gamma \in \Gamma$, we can efficiently
determine whether $\gamma$ contains all the points of $P$.
Alternatively, we want to report all the points of $P$ that lie
outside $\gamma$.
(This is clearly a special case of range emptiness searching or range
reporting, if one
considers the complements of the ranges in $\Gamma$.)
For simplicity, we only focus on the range fullness problem; the 
extension to reporting ``outliers'' is similar to the standard 
treatment of reporting queries, as discussed earlier.

We present a solution to this problem, with $O^*(n^{1/2})$
query time, thereby improving over the best known general bound
of $O^*(n^{2/3})$, given in \cite{AgM94}, which applies to any range
searching (e.g., range counting) with semi-algebraic sets 
(of constant description complexity) in $\reals^3$.

To apply our technique to this problem we first need to build a good
test set.
Since fullness searching is complementary to emptiness searching, we
need a property complementary to that assumed in 
Theorem~\ref{theo:part} (see also Lemma~\ref{lemma:const_cover}). 
In fact, we will enforce the property
that every full range $\gamma$ fully contains a single range of $Q$, 
which is ``almost full'' (contains at least $n-n/r$ points of $P$).

As above, assuming the ranges of $\Gamma$ to have $t$ degrees
of freedom, we map each range $\gamma\in\Gamma$ to a point 
$\gamma^*$ in $\reals^t$. 
A point $p\in\reals^3$ is mapped to a region $K_p$ which is the 
locus of all the points $\gamma^*$ that correspond to
ranges $\gamma$ which contain $p$.  We fix $r<n$,
take a random sample $N$ of $O(r\log r)$ points of $P$ 
(with a sufficiently large constant of proportionality), 
construct the {\em intersection} $I=\bigcap_{p\in N} K_p$, and 
decompose it into elementary cells. For each resulting cell 
$\sigma$, let $\gamma_\sigma$ denote the {\em intersection}
$\bigcap_{\gamma^*\in\sigma} \gamma$. As above, since $\sigma$ 
has constant description complexity, $\gamma_\sigma$ is a
semi-algebraic set of constant description complexity.
Note that, since the ranges in $\Gamma$ are convex, each range $\gamma_\sigma$
is also convex (albeit of potentially more complex shape than that of
the original ranges).

Define the test set $Q$ to consist of all the generalized 
ranges $\gamma_\sigma$, over all cells $\sigma$ in the decomposition
of $I$. We argue that $Q$ satisfies all four properties required from
a good test set:
(i) Compactness: As above, the size of $Q$ is polynomial in $r$ 
(it is at most $O^*(r^{2t-4})$).  
(ii) Shallowness (or, rather, ``almost fullness''):
For each cell $\sigma$ and any $\gamma\in\Gamma$ with
$\gamma^*\in\sigma$, $\gamma^*$ lies in all the sets $K_p$, 
for $p\in N$, and thus $N\subseteq \gamma$. By construction, we also
have $N\subseteq \gamma_\sigma$. Apply the $\eps$-net 
theory~\cite{HW} to the range space $(P,\tilde{\Gamma})$,
where the ranges of $\tilde{\Gamma}$ are complements of ranges of the
same form as the ranges $\gamma_\sigma$. 
Since $\gamma_\sigma^c\cap N=\emptyset$ for each cell $\sigma$ 
in the decomposition, we have, with high probability, the property 
that for each cell $\sigma$ of $I$,
$\gamma_\sigma$ contains at least $n-n/r$ points 
of $P$, so it is an ``almost full'' range.
(iii) Containment:
Let $\gamma\in\Gamma$ be a full range. Then, in particular, 
$N\subseteq \gamma$. Then $\gamma^*\in I$, and let $\sigma$ be the
cell of $I$ containing $\gamma^*$.  Then, by construction,
$\gamma_\sigma\subseteq \gamma$.
(iv) Efficiency: Finally, we show that
the complexity of a decomposition of the intersection of any $m$
ranges in $Q$, is $O^*(m^2)$, so $\zeta(m) = O^*(m^2)$.

\begin{claim}
\label{claim:decomposition_of_convex}
Let $Q$ be a set of convex ``almost full''
ranges, each containing at least $n - n/r$ points of $P$.
The intersection, $K$, of any $m$ ranges $q_1,\ldots,q_m \in Q$ 
can be decomposed into $O^*(m^2)$ elementary cells.
\end{claim}

\noindent{\it Proof.}
Since all ranges in $Q$ are convex, $K$ is a convex set too.
Assume, for simplicity of presentation, that $K$ is nonempty 
and has nonempty interior, and fix a point $o$ in that interior. 
We can regard the boundary of each $q_i$ as the graph of a 
bivariate function $\rho=F_i(\theta,\varphi)$ in spherical 
coordinates about $o$. Then $\bd K$ is the graph of the lower 
envelopes of these functions. Since the $q_i$'s have constant
description complexity, (the graph of) each $F_i$ is also a 
semi-algebraic set of constant description complexity\footnote{%
  With an appropriate algebraic re-parametrization of the spherical
  coordinates, of course.}.
Hence the combinatorial complexity of $\bd K$ is $O^*(m^2)$ \cite{SA}.
Moreover, since $\bd K$ is 2-dimensional, we can partition it into
$O^*(m^2)$ trapezoidal-like elementary cells, using a variant of
the vertical decomposition technique, and then extend each 
such cell $\tau_0$ to a 3-dimensional cone-like cell $\tau$, 
which is the union
of all segments connecting $o$ to the points of $\tau_0$. 
The resulting cells $\tau$ constitute a decomposition of $K$ 
into $O^*(m^2)$ elementary cells, as claimed.
$\Box$

Using the machinery developed in the preceding section,
we therefore obtain the following result.
\begin{theorem} \label{thm:full}
Let $P$ be a set of $n$ points in $\reals^3$, and let $\Gamma$ be a
family of convex ranges of constant description complexity. Then one
can construct, in near linear time, a data structure of linear 
size so that, for any range $\gamma\in\Gamma$, it can determine, 
in $O^*(n^{1/2})$ time, whether $\gamma$ is full.
\end{theorem}

\paragraph{Reporting outliers.} To extend the above approach to the problem
of reporting outliers, we apply a construction similar to that in the
``general recipe'' presented above. That is, we take the $b\log r$
deepest levels of $\A(N)$, for an appropriate constant $b$, decompose
them into elementary cells, and construct a generalized range
$\gamma_\sigma$ for each of these cells $\sigma$. The general machinery
given above implies the following result:

\begin{theorem} \label{thm:full_report}
Let $P$ be a set of $n$ points in $\reals^3$, and let $\Gamma$ be a
family of convex ranges of constant description complexity. Then one
can construct, in near linear time, a data structure of linear 
size so that, for any range $\gamma\in\Gamma$, it can report the points
of $P$ in the complement of $\gamma$,
in $O^*(n^{1/2}) + O(k)$ time, where $k$ is the query output size.
\end{theorem}

\subsection{Farthest point from a convex shape}

A useful application of the data structure of Theorem~\ref{thm:full}
is to {\em farthest point queries}.  In such a problem we are given 
a set $P$ of $n$ points in $\reals^3$, and wish to preprocess it, 
in near-linear time, into a data structure of linear size, 
so that, given a convex query object $o$ (from some fixed class of
objects with constant description complexity), we can efficiently
find the point of $P$ farthest from $o$.

We solve this problem using parametric searching \cite{Meg}. 
The corresponding decision problem is: 
Given the query object $o$ and a distance $\rho$,
determine whether the Minkowski sum $o\oplus B_\rho$ is full, where
$B_\rho$ is the ball of radius $\rho$ centered at the origin.
The smallest $\rho$ with this property is the distance to the 
farthest point from $o$. With an appropriate small-depth parallel
implementation of this decision problem, the parametric searching
also takes time $O^*(n^{1/2})$.
Reporting the $k$ farthest points from $o$, for any parameter $k$,
can be done in $O^*(n^{1/2}) + O(k)$ time, using a simple variant of
this technique.

\subsection{Computing the largest-area, largest-perimeter, and
largest-height triangles}

Let $P$ be a set of $n$ points in $\reals^d$. We wish
to find the triangle whose vertices belong to $P$ and 
whose area (respectively, perimeter, height) is maximal.
This problem is a useful subroutine in path approximation 
algorithms; see \cite{daescu}.
Daescu and Serfling \cite{daescu} gave an $O^*(n^{13/5})$-algorithm 
for the 3-dimensional largest-area triangle.
In $d$ dimensions, the running time is 
$O^*(n^{3-2/(\lfloor d^2/2 \rfloor + 1)})$.

In $\reals^3$, our technique, without any additional enhancements,
yields the improved bound $O^*(n^{5/2})$, using the 
following straightforward procedure. For each pair of points
$p_1,p_2 \in P$, we find the farthest point $q\in P$ from 
the line $\overline{p_1p_2}$, compute the area of $\Delta p_1p_2q$,
and output the largest-area triangle among those triangles. 
The procedure performs farthest-point queries from
$O(n^2)$ lines, for a total cost of $O^*(n^{5/2})$, as claimed.

We can improve this solution, using the following standard 
decomposition technique, to an algorithm with running
time $O^*(n^{26/11})$. First, the approach just described 
performs $M$ farthest-point queries on a set of $N$ points in 
time $O^*(MN^{1/2}+N)$, where the second term is the preprocessing 
cost of preparing the data structure.

Before continuing, we note the following technical issue. Recall that 
we find the farthest point from a query line $\ell$ by drawing a
cylinder $C_\rho$ around $\ell$, whose radius $\rho$ is the smallest
(unknown) radius for which $C_\rho$ contains $P$. The concrete value 
of $\rho$ is found using parametric searching. In the approach that we
follow now, we will execute in parallel $O(n^2)$ different queries, 
each with its own $\rho$, so care has to be taken when running the 
parametric search with this multitude of different unknown values 
of $\rho$.

While there are several alternative solutions to this problem, 
we use the following one, which seems the cleanest. Let $A>0$ be 
a fixed parameter. For each pair $p_1,p_2$ of distinct points of $P$, 
let $C_A(p_1p_2)$ denote the cylinder whose axis passes through 
$p_1$ and $p_2$ and whose radius is $2A/|p_1p_2|$. In the decision 
procedure, we specify the value of $A$, and perform $O(n^2)$ range 
fullness queries with the cylinders $C_A(p_1p_2)$. If all of them 
are found to be full, then $A\ge A^*$, where $A^*$ is the (unknown) 
maximal area of a triangle spanned by $P$; otherwise $A<A^*$. (With a
somewhat finer implementation, we can also distinguish between the
cases $A>A^*$ and $A=A^*$; we omit the details of this refinement.)

To implement the decision procedure, we apply a duality transform, 
where each cylinder $C$ in 3-space is mapped to a point 
$C^*=(a,b,c,d,\rho)$, where $(a,b,c,d)$ is some parametrization 
of the axis of $C$ and $\rho$ is its radius.  
In this dual parametric 5-space, a point $p\in\reals^3$ is
mapped to a surface $p^*$, which is the locus of all (points
representing) cylinders which contain $p$ on their boundary. Note that
the portion of space below (resp., above) $p^*$, in the 
$\rho$-direction, consists of points
dual to cylinders which do not contain (resp., contain) $p$.

Let $P^* = \{p^* \mid p\in P\}$. Fix some sufficiently large but
constant parameter $r_0$, and construct a $(1/r_0)$-cutting $\Xi$ 
of $\A(P^*)$, using the vertical decomposition of a random sample 
of $O(r_0\log r_0)$ surfaces of $P^*$ (see, e.g.,~\cite{SA}). 
As follows from \cite{CEGS,Kol}, the combinatorial complexity of 
$\Xi$ is $O^*(r_0^6)$.  We distribute the $O(n^2)$ points dual 
to the query cylinders among the cells of $\Xi$, in brute force, 
and also find, in equally brute force, for each cell $\tau$ the
subset $P^*_\tau$ of surfaces which cross $\tau$. We ignore cells
which fully lie below some surface of $P^*$, because cylinders whose
dual points fall in such a cell cannot be full (the decision algorithm
stops as soon as such a point (cylinder) is detected). For each of the
remaining cells $\tau$, we repeat this procedure with the subset of
the points dual to the surfaces in $P^*_\tau$ and with the subset of
cylinders whose dual points lie in $\tau$. 
We keep iterating in this manner until we reach cuttings whose cells
are crossed by at most $n/r$ dual surfaces, where $r$ is some
(non-constant) parameter that we will shortly fix. As is easily
checked, the overall number of cells in these cuttings is $O^*(r^6)$.

We then run the preceding weaker procedure on each of the resulting 
cells $\tau$, with the set $P_\tau$ of points dual to the surfaces 
which cross $\tau$ and with the set $\C_\tau$ of cylinders whose dual 
points lie in $\tau$. Letting $m_\tau$ denote the number of these 
cylinders, the overall cost of the second phase of the procedure is
$$
\sum_\tau O^*(m_\tau (n/r)^{1/2} + n/r) = O^*(n^2(n/r)^{1/2} + nr^5).
$$
Since $r_0$ is a constant, the overall cost of the first phase is
easily seen to be proportional to the overall size of the resulting
subproblems, which is $O^*(n^2+nr^5)$. Overall, the cost is thus
$$
O^*(n^{5/2}/r^{1/2} + nr^5).
$$
Choosing $r=n^{3/11}$, this becomes $O^*(n^{26/11})$.

Running a generic version of this decision procedure in parallel is
fairly straightforward. The cuttings themselves depend only on the
dual surfaces, which do not depend on $A^*$, so we can construct them
in a concrete, non-parametric fashion. Locating the points dual to the
query cylinders can be done in parallel, and, since $r_0$ is a
constant, this takes constant parallel depth for each of the
logarithmically many levels of cuttings. The second phase can also be
executed in parallel in an obvious manner. Omitting the further easy
details, we conclude that the overall algorithm also takes
$O^*(n^{26/11})$ time.

\paragraph{Largest-perimeter triangle.}
The above technique can be adapted to yield efficient solutions of
several problems of a similar flavor. For example, consider the
problem of computing the largest-perimeter triangle among those 
spanned by a set $P$ of $n$ points in $\reals^3$. 
Here, for each pair $p_1,p_2$ of points of $P$, and for a specified
perimeter $\pi$, we construct the ellipsoid of revolution
$E_\pi(p_1,p_2)$, whose boundary is the locus of all points $q$
satisfying $|qp_1|+|qp_2| = \pi-|p_1p_2|$. (Here, of course, we only
consider pairs $p_1,p_2$ with $|p_1p_2| < \pi/2$.)
We now run $O(n^2)$ range fullness queries with these ellipsoids, and
report that $\pi^*>\pi$ if at least one of these ellipsoids in not
full, or $\pi^*\le\pi$ otherwise, where $\pi^*$ is the largest
perimeter.

The efficient implementation of this procedure is carried out similar
to the preceding algorithm, except that here the dual representation
of our ellipsoids require six degrees of freedom, to specify the
foci $p_1$ and $p_2$.
Unlike the previous case, the dual surfaces $p^*$ do depend on $\pi$,
so, in the generic implementation of the decision procedure we also
need to construct the various $(1/r_0)$-cuttings in a generic,
parallel manner.\footnote{%
  We can make these surfaces independent of $\pi$ if we add
  $\pi$ as a seventh degree of freedom, but then the overall performance
  of the algorithm deteriorates.}
However, since $r_0$ is a constant, this is easy to do in constant
parallel depth per cutting.
A $(1/r_0)$-cutting in $\reals^6$ has complexity
$O^*(r^{8})$ \cite{CEGS,Kol}. A modified version of the preceding
analysis then yields:
\begin{theorem}
The largest-perimeter triangle among those spanned by a set of $n$ 
points in $\reals^3$ can be computed in $O^*(n^{12/5})$ time.
\end{theorem}

\paragraph{Largest-height triangle.}
In this variant, we wish to compute the triangle with largest height
among those determined by a set $P$ of $n$ points in $\reals^3$.
Here, for each pair $p_1,p_2$ of points of $P$, and for a specified
height $h$, we construct the cylinder $C_h(p_1,p_2)$, whose axis 
passes through $p_1$ and $p_2$ and whose radius is $h$. 
We run $O(n^2)$ range fullness queries with these cylinders, and
report that $h^*>h$ if at least one of these cylinders in not
full, or $h^*\le h$ otherwise, where $h^*$ is the desired
largest height.

The efficient implementation of this procedure is carried out as
above, except that here the dual representation of these cylinders
require only four degrees of freedom, once $h$ is specified.
As in the preceding case, here too the surfaces of $P^*$ also 
depend on $h$, so we need a generic parallel procedure for 
constructing $(1/r_0)$-cuttings for these surfaces,
which however is not difficult to achieve, since
$r_0$ is a constant. We omit the simple routine details.
Since a $(1/r_0)$-cutting in $\reals^4$ has complexity
$O^*(r^{4})$ \cite{Kol}, a modified version of the preceding
analysis then yields:
\begin{theorem}
The largest-height triangle among those spanned by a set of $n$ 
points in $\reals^3$ can be computed in $O^*(n^{16/7})$ time.
\end{theorem}

\paragraph{Further extensions.}
We can extend this machinery to higher dimensions, although its
performance deteriorates as the dimension grows. The range fullness
problem in $\reals^d$, for $d\ge 4$, can be handled in much the same
way as in the 3-dimensional case. When extending
Claim~\ref{claim:decomposition_of_convex},
we have an intersection of $m$ convex sets of constant description
complexity in $\reals^d$, and we can regard the boundary of the
intersection as the lower envelope of $m$ $(d-1)$-variate functions of
constant description complexity, each representing the boundary of one
of the input convex sets, in spherical coordinates about some fixed
point in the intersection. The complexity of the lower envelope is
$O^*(m^{d-1})$ \cite{Sh}. However, we need to decompose the region
below the envelope into elementary cells, and, as already noted, the
only known general-purpose technique for doing so is to decompose the
entire arrangement of the graphs of the $m$ boundary functions, and
select the cells below the lower envelope. The complexity of such a
decomposition is $O^*(m^{2d-4})$ \cite{CEGS,Kol}. This implies that
$\zeta(r)=O^*(r^{2d-4})$. The rest of the analysis, including the
construction of a good test set, is done in essentially the same
manner. Hence, using the machinery of the previous section, we obtain:

\begin{theorem} \label{fulld}
Let $P$ be a set of $n$ points in $\reals^d$, for $d\ge 4$, and let
$\Gamma$ be a family of convex ranges of constant description
complexity. Then one can construct, in near linear time, a data
structure of linear size so that, for any range $\gamma\in\Gamma$, 
it can determine, in $O^*(n^{1-1/(2d-4)})$, whether $\gamma$ is full.
\end{theorem}

\paragraph{Finding the largest-area triangle in $\reals^d$.}
Let $P$ be a set of $n$ points in $\reals^d$, for $d\ge 4$, and
consider the problem of finding the largest-area triangle spanned by
$P$. We apply the same method as in the 3-dimensional case, whose
main component is a decision procedure which tests $O(n^2)$ cylinders
for fullness. A cylinder (with a line as an axis) in $\reals^d$ has
$2d-1$ degrees of freedom, so the dual representation of our $O(n^2)$
cylinders is as points in $\reals^{2d-1}$. The best known bound on
the complexity of a $(1/r)$-cutting in this space is
$O^*(r^{2(2d-1)-4}) = O^*(r^{4d-6})$. Applying this bound and the
bound in Theorem~\ref{fulld}, the overall cost of the decision
procedure is
$$
O^*\left( n^2(n/r)^{1-1/(2d-4)} + nr^{4d-7} \right) .
$$
Optimizing the value of $r$, and applying parametric searching,
we get an algorithm for the maximum-area triangle in $\reals^d$
with running time 
$$
O^*\left( n^{1+\frac{(4d-9)(4d-7)}{(4d-6)(2d-4)-1}} \right) .
$$
We can extend the other problems (largest-perimeter or largest-height
triangles) in a similar manner, and can also obtain algorithms for
solving higher-dimensional variants, such as computing the
largest-volume tetrahedron or higher-dimensional simplices. We omit
the straightforward but tedious analysis, and the resulting
cumbersome-looking bounds.

\section{Ray shooting amid balls in 3-space}
\label{sec:balls}

Let $\B$ be a set of $n$ balls in 3-space. We show how to preprocess 
$\B$ in near-linear time into a data structure of linear size, so
that, given a query ray $\rho$, the first ball that $\rho$ hits
can be computed in $O^*(n^{2/3})$ time, improving the general bound
$O^*(n^{3/4})$ mentioned in the introduction.
As already noted, we use the parametric-searching technique
of Agarwal and Matou\v{s}ek~\cite{AgM93}, which reduces the problem
to that of efficiently testing whether a query segment 
$s = qz \subset \rho$ intersects any ball in $\B$, where $q$ is the
origin of $\rho$ and $z$ is a parametric point along $\rho$.

\paragraph{Parametric representation of balls and segments.}
We move to a parametric 4-dimensional space, in which balls 
in 3-space are represented by points, so that a ball with center
at $(a,b,c)$ and radius $r$ is mapped to the point $(a,b,c,r)$.
A segment $e$, or for that matter, any closed nonempty
set $K\subset \reals^3$ of constant description complexity,
is mapped to a surface $\sigma_K$, which
is the locus of all points representing balls that touch $K$ but are
openly disjoint from $K$. By construction, $\sigma_K$ is the graph of
a totally defined continuous trivariate function $r=\sigma_K(a,b,c)$,
which is semi-algebraic of constant description complexity. Moreover,
points below (resp., above) $\sigma_K$ represent balls which are
disjoint from $K$ (resp., intersect $K$). 

Moreover, for any such set $K$, $\sigma_K(q)$ is, by definition,
the (Euclidean) distance of $q$ from $K$. Hence, given a collection
$\K=\{K_1,K_2,\ldots,K_m\}$ of $m$ sets, the minimization 
diagram of the surfaces $\sigma_{K_1},\ldots,\sigma_{K_m}$ 
(that is, the projection onto the 3-space $r=0$ of the lower 
envelope of these surfaces) is the nearest-neighbor Voronoi 
diagram of $\K$. We use this property later on, in deriving a
sharp bound on the resulting function $\zeta(\cdot)$.

\paragraph{Building a test set for segment emptiness.}
Here we use the general recipe for constructing good test sets, which
covers each empty segment $e$ by a fairly complex ``canonical''
empty region $K$, which has nonetheless constant description
complexity. In parametric 4-space, each such region $K$ is mapped to 
the upper halfspace above the corresponding surface $\sigma_K$; 
this is the set of all balls that intersect $K$. 
The complement of the union of $m$
such ranges is the portion of 4-space below the lower envelope of the
corresponding surfaces $\sigma_{K_i}$. Using the connection between
this envelope and the Voronoi diagram of the $K_i$'s, we are able to
decompose (the diagram and thus) the complement of the union into
$\zeta(m)=O^*(m^3)$ elementary cells.

in more detail, the construction proceeds as follows.
We start by choosing a random sample $N$ of $O(r\log r)$ balls of $\B$,
to construct
a test set $Q$ for empty segment ranges, with respect to $N$.
While we do not have a clean, explicit geometric definition of 
these ranges, they will satisfy, as above, all the four
requirements from a good test set. Also, we spell out the
adaptation of the general recipe to the present scenario, to help the
reader see through one concrete application of the general recipe.

Specifically, we move to a dual space, in which segments in 3-space
are represented as points. Segments in 3-space have six degrees
of freedom; for example, we can represent a segment by the 
coordinates of its two endpoints.
The dual space is therefore 6-dimensional. 
Each ball $B\in N$ is mapped to a surface $B^*$, which is the 
locus of all points representing segments which touch $\bd B$ but do
not penetrate into its interior; that is, either they are tangent to 
$B$, at a point in their relative interior, or they have an endpoint 
on $\bd B$ but are openly disjoint from $B$. 

Let $N^*$ denote the collection of the surfaces dual to the balls of
$N$. Construct a $(1/r)$-cutting of $\A(N^*)$, which consists of
$O^*(r^8)$ elementary cells \cite{CEGS,Kol}. Each cell $\tau$ has the
property that all points in $\tau$ represent segments which meet a
fixed set of balls from among the balls in $N$ and avoid all other
balls of $N$; the set depends only on $\tau$. 

For each cell $\tau$ whose corresponding set of balls is
\emph{empty}, we define $K_\tau$ to be the union, in 3-space,
of all segments $e$ whose dual points lie in $\tau$. 
Since $\tau$ is an elementary cell, $K_\tau$ is a semi-algebraic 
set of constant description complexity (see, e.g., \cite{BPR06}).
Moreover, $K_\tau$ is an {\em $N$-empty} region, in the 
sense that it is openly disjoint from all the balls in $N$. 

Since we have to work in parametric 4-space, we map each region
$K_\tau$ of the above kind into a range $\gamma_\tau$ in 4-space, 
which is the locus of all (points representing) balls which intersect
$K_\tau$. As discussed above, $\gamma_\tau$ is the upper halfspace 
bounded by the graph $\sigma_{K_\tau}$ of the distance function from 
points in $\reals^3$ to $K_\tau$.

We define the desired test set $Q$ to consist of all the ranges
$\gamma_\tau$, as just defined, and argue that $Q$ indeed satisfies
all four properties required from a good test set:
(a) Compactness: $|Q|=O^*(r^8)$, so its size is small. 
(b) Shallowness: With high probability, each range in $Q$ is 
$(n/r)$-shallow, since it does not contain any point 
representing a ball in $N$ (and we assume that the sample 
$N$ does indeed have this property, which
makes all the ranges in $Q$ $(n/r)$-shallow).
(c) Containment:
Each empty segment is also $N$-empty, so its dual point lies 
in some cell $\tau$ of the cutting, whose associated subset of balls
is empty. By construction, we have $e\subset K_\tau$. 
That is, any ball intersecting $e$ also intersects $K_\tau$, so
the range in 4-space that $e$ defines is contained in $\gamma_\tau$,
i.e., in a single range of $Q$.
(d) Efficiency:
The complement of the union of any $m$ ranges in $Q$ can be
decomposed into $O^*(m^3)$ elementary cells.

The proof of (d) proceeds as follows. 
The complement of the union of $m$ ranges,
$\gamma_{\tau_1},\ldots,\gamma_{\tau_m}$, is the region 
below the lower envelope of the corresponding surfaces
$\sigma_{K_{\tau_1}},\ldots,\sigma_{K_{\tau_m}}$.
To decompose this region, it suffices to produce
a decomposition of the 3-dimensional minimization diagram of these
surfaces, and extend each of the resulting cells into a 
semi-unbounded vertical prism, whose ``ceiling'' lies 
on the envelope.  

The combinatorial complexity of the minimization diagram of a 
collection $\K=\{K_{\tau_1},\ldots,K_{\tau_m}\}$ of $m$ trivariate 
functions of constant description complexity is\footnote{%
  This bound holds regardless of how ``badly'' the regions in $\K$ 
  are shaped, nor how ``wildly'' they can intersect one another,
  as long as each of them has constant description complexity.}
$O^*(m^3)$~\cite{SA}.
Moreover, as noted above, the minimization diagram is the
Euclidean nearest-neighbor Voronoi diagram of $\K$.

We can decompose each cell $V_i = V(K_{\tau_i})$ of the diagram
(or, more precisely,
the portion of the cell outside the union of the $K_{\tau_i}$'s) using its 
{\em star-shapedness} with respect to its ``site'' $K_{\tau_i}$;
that is, for any point $p \in V(K_{\tau_i})$, the segment connecting
$p$ to its nearest point on $K_{\tau_i}$ is fully contained in
$V(K_{\tau_i})$.
As is easy to verify, this property holds regardless of the shape, or intersection pattern,
of the regions in $\K$. We
first decompose the 2-dimensional faces bounding $V_i$ into 
elementary cells,
using, e.g., an appropriate variant of 2-dimensional vertical decomposition,
and then take each such cell $\phi_0$ and extend it
to a cell $\phi$, which is the union of all segments, each connecting 
a point in $\phi_0$ to its nearest point on $K_{\tau_i}$. The
resulting cells, obtained by applying this decomposition to all 
cells of the diagram, form a decomposition of the portion of the 
diagram outside the union of the $K_{\tau_i}$'s, into a total of
$O^*(m^3)$ elementary cells, as desired. The union of the
$K_{\tau_i}$'s themselves, being a subcollection of cells
of a 3-dimensional arrangement of $m$ regions of constant description
complexity, can also be decomposed into $O^*(m^3)$
cells, using standard results on vertical decomposition in three
dimensions~\cite{SA}.

Using Lemma~\ref{lemma:const_cover} and the machinery of 
Section~\ref{section:range_emptiness}, in conjunction with the
parametric searching technique of \cite{AgM93},
we thus obtain the following theorem.
\begin{theorem} \label{main_balls}
Ray shooting amid $n$ balls in 3-space can be performed in
$O^*(n^{2/3})$ time, using a data structure of
$O(n)$ size, which can be constructed in $O^*(n)$ time.
\end{theorem}

\paragraph{Remark:} In the preceding description, we only considered
{\em empty} ranges. If desired, we can extend the analysis to obtain a data
structure which also supports ``reporting queries'', in which we want
to report the first $k$ balls hit by a query ray. We omit the details of
this straightforward extension.

\section{Range emptiness searching and reporting in the plane} 

\paragraph{Fat triangle reporting and emptiness searching.}
Let $\alpha > 0$ be a fixed constant, and let $P$ be a set of 
$n$ points in the plane. We wish to preprocess $P$, in $O^*(n)$ time,
into a data structure of size $O(n)$, which, given
an $\alpha$-fat query triangle $\gamma$ (which, as we recall, is a
triangle all of whose angles are at least $\alpha$), 
can determine in $O^*(1)$ time whether $\gamma\cap P=\emptyset$, 
or report in $O^*(1) + O(k)$ time
the points of $P$ in $\gamma$, where $k = |P \cap \gamma|$.

\begin{figure}[htbp]
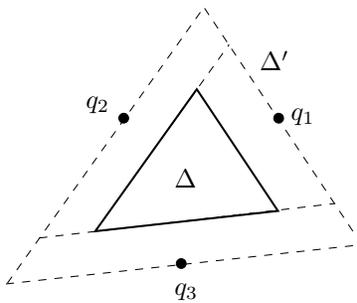

\begin{center}
\input tricanon1.pstex_t
\caption{The first step in canonizing an empty triangle.}
\label{fig:tricanon1}
\end{center}
\end{figure}

To do so, we need to construct a good test set $Q$.
We use the following ``canonization''
process (an ad-hoc process, not following the general recipe of
Section~\ref{section:range_emptiness}).
As above, we apply the construction 
to a random sample $N$ of $O(r\log r)$ points of $P$.
For simplicity, we first show how to canonize {\em empty} triangles,
and then extend the construction to shallow triangles.
(As before, the first part suffices for emptiness searching,
whereas the second part is needed for reporting queries.)
Let $\Delta$ be an $\alpha$-fat empty triangle, which is then also
$N$-empty. We expand $\Delta$ homothetically, keeping one vertex 
fixed and translating the opposite side away, until it hits a point 
$q_1$ of $N$. We then expand the new triangle homothetically from 
a second vertex, until the opposite side hits a second point $q_2$ 
of $N$, and then apply a similar expansion from the third vertex, 
making the third edge of the triangle touch a third point $q_3$ of 
$N$. We end up with an $N$-empty triangle $\Delta'$, homothetic to,
and containing, $\Delta$, each of whose sides passes through one 
of the points $q_1,q_2,q_3\in N$. See Figure~\ref{fig:tricanon1}.
(It is possible that some of these expansions never hit a point 
of $N$, so we may end up with an unbounded wedge or halfplane 
instead of a triangle. Also, the points $q_1,q_2,q_3$ need not be
distinct.)

\begin{figure}[htbp]
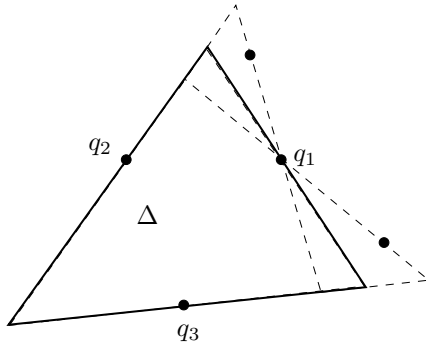

\begin{center}
\input tricanon2.pstex_t
\caption{The second step in canonizing an empty triangle.}
\label{fig:tricanon2}
\end{center}
\end{figure}

Let $\D$ be the set of orientations 
$\{j\alpha /4 \mid j=0,1,\ldots,\lfloor 8\pi/\alpha\rfloor \}$.
We turn the side containing $q_1$ clockwise and counterclockwise
about $q_1$, keeping its endpoints on the lines containing the other
two sides, until we reach an orientation in $\D$, or until we 
hit another point of $N$ (which could also be one of the points 
$q_2,q_3$), whichever comes first. Each of the new sides forms, with
the two lines containing the two other sides, a new (openly)
$N$-empty $(3\alpha/4)$-fat triangle; the union of these two triangles
covers $\Delta'$. See Figure~\ref{fig:tricanon2}.

For each of the two new triangles, $\Delta''$, we apply the same
construction, by rotating the side containing $q_2$ clockwise 
and counterclockwise, thereby obtaining two new triangles
whose union covers $\Delta''$. We then apply the same construction to
each of the four new triangles, this time rotating about $q_3$.
Overall, we get up to eight new triangles whose union covers $\Delta$. 
Each of these new triangles is $(\alpha/2)$-fat, openly $N$-empty, 
and each of its sides either passes through two points of $N$, 
or passes through one point of $N$ and has orientation in $\D$. 
Since $|\D|=O(1/\alpha)=O(1)$, it follows that the overall number of 
these canonical covering triangles is $O((r\log r)^6)=O^*(r^6)$.
(We omit the easy extensions of this step to handle unbounded wedges
or halfplanes, or the cases where the points $q_i$, or some of the
newly encountered points, lie at vertices of the respective triangles.)

We take $Q$ to be the collection of these canonical triangles, and
argue that $Q$ indeed satisfies the properties of a good test set:
(a) Compactness: $|Q|=O^*(r^6)$, so its size is small. 
(b) Shallowness: With high probability, each range in $Q$ is 
$(n/r)$-shallow (and, as usual, we assume that this property 
does indeed hold).
(c) Containment: By construction, each $\alpha$-fat empty 
triangle is contained in the union of at most eight triangles in $Q$.
(d) Efficiency: 
Being $(\alpha/2)$-fat, the union of any $m$ triangles in $Q$ has
complexity $O(m\log\log m)$ \cite{MPSSW}, so the associated function
$\zeta$ satisfies $\zeta(m)=O(m\log\log m)$.
This, combined with Lemma~\ref{lemma:const_cover} and the 
machinery of Section~\ref{section:range_emptiness}, 
lead to the following theorem.

\begin{theorem} \label{theorem:fat_triangles_emptiness}
One can preprocess a set $P$ of $n$ points in the plane, in 
near-linear time, into a data structure of linear size, 
so that, for any query $\alpha$-fat triangle $\Delta$, one can 
determine, in $O^*(1)$ time, whether $\Delta\cap P=\emptyset$.
\end{theorem}

\paragraph{Reporting points in fat triangles.}
We can extend the technique given above to the problem of reporting
the points of $P$ that lie inside any query fat triangle. For this, we
need to construct a test set that will be good for shallow ranges and
not just for empty ones. Using Theorem~\ref{epsshallow}, we construct 
(by random sampling) a shallow $(1/r)$-net $N\subseteq P$ of size 
$O(r\log r)$. We next canonize every $(n/r)$-shallow $\alpha$-fat
triangle $\Delta$, by the same canonization process used above, with
respect to the set $N$. Note that each of the resulting canonical
triangles contains (in its interior)
the same subset of $N$ as $\Delta$ does. By the
properties of shallow $(1/r)$-nets, since $|\Delta\cap P|\le n/r$,
we have $|\Delta\cap N| = O(\log r)$, so all the resulting canonical
triangles are $(c\log r)$-shallow with respect to $N$, for some
absolute constant $c$. Again, since $N$ is a shallow $(1/r)$-net,
all the canonical triangles are $(c'n/r)$-shallow with respect to $P$,
for another absolute constant $c'$. Hence, the resulting collection 
$Q$ of canonical triangles is a good test set for all shallow fat 
triangles, and we can apply the machinery of
Section~\ref{section:range_emptiness} to
obtain a data structure of linear size, which can be constructed in
near-linear time, and which can perform reporting queries in fat
triangles in time $O^*(1) + O(k)$, where $k$ is the output size of the
query.

\paragraph{Range emptiness searching with semidisks and circular
caps.}
The motivation for studying this problem comes from the following
problem, addressed in \cite{DMSW}. We are given a set $P$ of $n$
points in the plane, and wish to preprocess it into a data structure
of linear size, so that, given a query point $q$ and a query line
$\ell$, one can quickly find the point of $P$ closest to $q$ and lying
above $\ell$. In the original problem, as formulated in \cite{DMSW},
one also assumes that $q$ lies on $\ell$, but we will consider, and
solve, the more general version of the problem, where $q$ also lies 
above $\ell$.

The standard approach (e.g., as in \cite{AgM94}) yields a solution
with linear storage and near-linear preprocessing, and query time
$O^*(n^{1/2})$. We present a solution with query time $O^*(1)$.

Using parametric searching \cite{Meg}, 
the problem reduces to that of testing
whether the intersection of a disk of radius $\rho$ centered at $q$
with the halfplane $\ell^+$ above $\ell$ is $P$-empty. The resulting 
range is a circular cap larger than a semidisk (or exactly a 
semidisk if $q$ lies on $\ell$).
Again, the main task is to construct a good test set $Q$ for such
ranges, which we do by using an ad-hoc canonization process, 
which covers each empty circular cap by $O(1)$ canonical caps,
which satisfy the properties of a good test set; in particular,
we will have $\zeta(m) = O^*(m)$.
(As before, we consider here only the case of emptiness detection,
and will consider the reporting problem later.)

\begin{figure}[htbp]
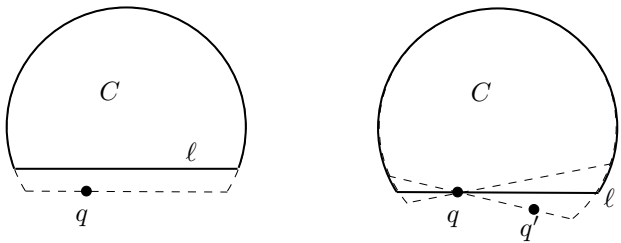

\begin{center}
\input capcanon1.pstex_t
\caption{The first steps in canonizing an empty cap.}
\label{fig:capcanon1}
\end{center}
\end{figure}

To construct a test set $Q$ we choose a random sample $N$ of $O(r\log r)$
points of $P$ and build a set of canonical empty ranges with respect to $N$.
Let $C=C_{c,\rho,\ell}$ be a given circular cap (larger than a 
semidisk) with center $c$, radius $\rho$, and chord supported 
by a line $\ell$. We first translate $\ell$ in the direction which
enlarges the cap, until either its portion within the disk $D$ of 
the cap touches a point of $N$, or $\ell$ leaves $D$.
See Figure~\ref{fig:capcanon1}(left).
In the latter case, $C$ is contained in a complete
$N$-empty disk, and it is fairly easy to show that such a disk is
contained in the union of at most three canonical $N$-empty disks,
each passing through three points of $N$ or through two diametrically
opposite points of $N$; there are at most $O^*(r^3)$ such disks.

Suppose then that the new chord (we continue to denote its line 
as $\ell$) passes through a point $q$ of $N$, as in the figure. 
Let $\D$ be a set of $O(1)$
canonical orientations, uniformly spaced and sufficiently dense
along the unit circle, for some small constant value $\alpha$.
Rotate $\ell$ about $q$ in both clockwise and counterclockwise 
directions, until we reach one of the two following events: 
(i) the orientation of $\ell$ belongs to $\D$; or 
(ii) the portion of $\ell$ within $D$ touches another point of $N$.
In either case, the two new lines, call them $\ell_1,\ell_2$,
become canonical---there are only $O^*(r^2)$ such possible lines.
Note that our original cap $C$ is contained in the union
$C_1\cup C_2$, where $C_1=C_{c,\rho,\ell_1}$ and 
$C_2=C_{c,\rho,\ell_2}$. Moreover, although the
new caps need no longer be larger than a semidisk, they 
are not much smaller---this is an easy exercise in elementary 
geometry. See Figure~\ref{fig:capcanon1}(right).

\begin{figure}[htbp]
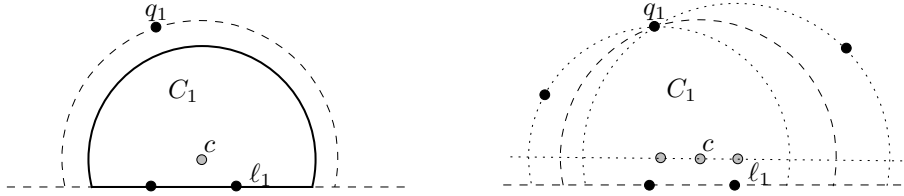

\begin{center}
\input capcanon2.pstex_t
\caption{The second step in canonizing an empty cap.}
\label{fig:capcanon2}
\end{center}
\end{figure}

We next canonize the disk of $C$ (which is also the disk of $C_1$ 
and $C_2$). Fix one of the new caps, say $C_1$. Expand $C_1$ from 
the center $c$, keeping the line $\ell_1$ fixed, until we hit a point $q_1$ 
of $N$ (lying in $\ell_1^+$). See Figure~\ref{fig:capcanon2}(left).
If $c$ lies in $\ell_1^+$ then 
we move $c$ parallel to $\ell_1$ in both directions, 
again keeping $\ell_1$ itself fixed and keeping the 
circle pass through $q_1$, until we obtain two circular caps, 
each passing through $q_1$ and through a second point of $N$ 
(if we do not hit a second point, we reach a quadrant, bounded 
by $\ell$ and by the line orthogonal to $\ell$ through $q_1$). 
The union of the two new circular caps covers $C_1$. 
See Figure~\ref{fig:capcanon2}(right).

\begin{figure}
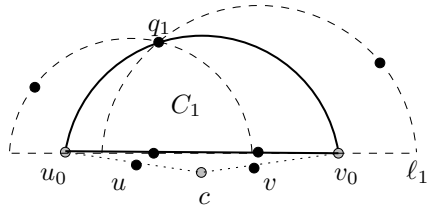

\begin{center}
\input c1lemma.pstex_t
\caption{$C_1$ is contained in the union of the two other caps.}
\label{fig:c1lemma}
\end{center}
\end{figure}

If $c$ lies in $\ell_1^-$, we move it along the two rays 
connecting it to the endpoints $u_0,v_0$ of the chord defined 
by $\ell_1$. As before, each of the motions stops when the 
circle hits another point of $N$ in $\ell_1^+$, or when the motion 
reaches $u_0$ or $v_0$. We claim that $C_1$ is
contained in the union of the two resulting caps. 
Indeed, let $u$ and $v$ denote the locations of the center 
at the two stopping placements. We need to show that, for any point
$b\in C_1$ we have either $|bu| \le |q_1u|$
or $|bv| \le |q_1v|$. If both inequalities did not hold, 
then both $u$ and $v$ would have to lie on the side of the 
perpendicular bisector of $q_1b$ containing $q_1$. This is easily seen
to imply that $c$ must also lie on that side, which however is
impossible (because $|bc|\le|q_1c|$).
See Figure~\ref{fig:c1lemma}.

Next, we take one of these latter caps, $C'$, whose bounding
circle passes through $q_1$ and through a second point $q_2$ of 
$N\cap \ell_1^+$, and move its center along the bisector of 
$q_1q_2$ in both directions, keeping the bounding circle touch 
$q_1$ and $q_2$, and still keeping the line $\ell_1$ supporting 
the chord fixed. We stop when the first of these events takes place:
(i) The center reaches $\ell_1$, in which case the cap becomes a
semidisk (this can happen in only one of the moving directions).
(ii) The center reaches the midpoint of $q_1q_2$.
(iii) The bounding circle touches a third point of $N\cap\ell_1^+$.
(iv) The central angle of the chord along $\ell_1$ is equal to
some fixed positive angle $\beta>0$.
The union of the two new caps covers $C'$.
(It is possible that during the motion the moving circle becomes 
tangent to $\ell_1$, and then leaves it, in which case the 
corresponding final cap is a full disk.)

Similarly, if the center of $C'$ lies on $\ell_1$ (which can
happen when the motion in the preceding canonization step reaches
$v_0$ or $u_0$), then we canonize its disk by translating the center
to the left and to the right along $l_1$ until the bounding circle
touches another point of $N\cap\ell_1^+$, exactly as in the preceding case
(shown in Figure~\ref{fig:capcanon2}(right)).

Let $C''$ be one of the four new caps. In all cases $C''$ is
canonical: For the first kind of caps,
the stopping condition that defines $C''$ is (ii) 
or (iii) then either the circle bounding $C''$ passes through 
three points of $N$ or it passes through two diametrically 
opposite points of $N$. There are a total of $O^*(r^3)$ such 
circles, and since $C''$ is obtained (in a unique manner)
by the interaction of one of these circles and one of the 
$O^*(r^2)$ canonical chord-lines, there is a total of $O^*(r^5)$ 
such caps.  If the stopping condition is (i), the cap is also
canonical, because the center of the containing disk is the
intersection point of a bisector of two points of $N$ with
one of the $O^*(r^2)$ canonical chord-lines, so there is a total 
of $O^*(r^4)$ such caps.
In the case of condition (iv), there are only $O^*(r^2)$ such
disks, for a total of  $O^*(r^4)$ caps. Similar reasoning  shows that
the caps resulting in the second case are also canonical, and their number
is $O^*(r^4)$.

We take the test set $Q$ to consist of all the caps of the 
final forms, and argue that it satisfies the properties of a 
good test set:
(a) Compactness: $|Q|=O^*(r^5)$, so its size is small. 
(b) Shallowness: 
With high probability, each range in $Q$ is $(n/r)$-shallow
(and we assume that this property does indeed hold).
(c) Containment:
Each empty cap is also $N$-empty, so, by the above canonization
process, it is contained in the union of $O(1)$ caps of $Q$.
(d) Efficiency: Each cap $C\in Q$ is {\em $(\alpha,\beta)$-covered}, 
for appropriate fixed constants $\alpha,\beta>0$, in the 
terminology of \cite{Ef05}, meaning that for each point 
$p\in\bd C$ there exists an $\alpha$-fat triangle touching $p$, 
contained in $C$, and with diameter which is 
at least $\beta$ times the diameter of $C$. In addition, the
boundaries of any two ranges in $Q$ (or of any two circular caps, for
that matter) intersect in at most four points, as is easily checked.
As follows from the recent analysis of de Berg \cite{dB08}, 
the complexity of the union of any $m$ ranges in $Q$ is 
$O(\lambda_6(m)\log^2m) = O^*(m)$. Hence, the complement of 
the union of any $m$ ranges in $Q$ can be
decomposed into $O^*(m)$ elementary cells, 
making $\zeta(m) = O^*(m)$.

In conclusion, we obtain:

\begin{theorem} \label{thm:caps}
Let $P$ be a set of $n$ points in the plane. We can preprocess $P$, in
near-linear time, into a data structure of linear size, so that, for
any query circular cap $C$, larger than a semidisk, we can test
whether $C\cap P$ is empty, in $O^*(1)$ time.
\end{theorem} 

Combining Theorem~\ref{thm:caps} with parametric searching, we obtain:

\begin{corollary} \label{cor:caps}
Let $P$ be a set of $n$ points in the plane. We can preprocess $P$, in
near-linear time, into a data structure of linear size, so that, for
any query halfplane $\ell^+$ and point $q\in\ell^+$, we can find, 
in $O^*(1)$ time, the point in $P\cap\ell^+$ nearest to $q$.
\end{corollary} 

\noindent{\bf Remark.}
The machinery developed in this section also
applies to smaller circular caps, as long as they are not too small.
Formally, if the central angle of each cap is at least some fixed
constant $\alpha>0$, the same technique holds, so we can test
emptiness of such ranges in $O^*(1)$ time, using a data structure
which requires $O(n)$ storage and $O^*(n)$ preprocessing.
Thus Theorem~\ref{thm:caps} carries over to this scenario, but 
Corollary \ref{cor:caps} does not, because we have no control over 
the ``fatness'' of the cap, as the disk shrinks or expands, when 
the center of the disk lies in $\ell^-$, and once the canonical caps
become too thin, the complexity of their union may become quadratic.

\paragraph{Reporting points in semidisks and circular caps.}
As in the case of fat triangles, we can extend the technique to answer
efficiently range reporting queries in semidisks or in sufficiently
large circular caps. We use the same canonization process, with
respect to a random sample $N$ of size $O(r\log r)$ which is a shallow
$(1/r)$-net, and argue, exactly as in the case of fat triangles, that
the resulting collection of canonical caps is a good test set for
shallow semidisk or larger cap ranges. Applying the machinery of
Section~\ref{section:range_emptiness},
we obtain a data structure of linear size, 
which can be constructed in near-linear time, and which can 
perform reporting queries in semidisks or larger caps, in time 
$O^*(1) + O(k)$, where $k$ is the output size of the query.


\section{Approximate range counting}
\label{sec:apx}

Given a set $P$ of $n$ points in $\reals^d$, a set $\Gamma$ of 
semi-algebraic ranges of constant description complexity, and 
a parameter $\delta>0$, the {\em approximate range counting} 
problem is to preprocess $P$ into a data structure such that, 
for any query range $\gamma \in \Gamma$, we can efficiently 
compute an approximate count $t_\gamma$ which satisfies
$$
(1-\delta)|P\cap\gamma| \le t_\gamma \le (1+\delta)|P\cap\gamma| .
$$
As in most of the rest of the paper, we will only consider the case
where the size of the data structure is to be (almost) linear, 
and the goal is to find solutions with small query time.

The problem has been studied in several recent
papers~\cite{AHP,AHPS,ArS,KRS}, for the special case where $P$ 
is a set of points in $\reals^d$ and $\Gamma$ is the collection 
of halfspaces (bounded by hyperplanes). A variety of solutions, 
with near-linear storage, were derived; in all of them, the 
dependence of the query cost on $n$ is close to 
$n^{1-1/\lfloor d/2\rfloor}$, which, as reviewed earlier, is 
roughly the same as the cost of halfspace range emptiness queries, 
or the overhead cost of halfspace range reporting 
queries~\cite{Ma:rph}.

The fact that the approximate range counting problem is closely
related to range emptiness comes as no surprise, because, when
$P\cap\gamma=\emptyset$, the approximate count $t$ must be $0$, so
range emptiness is a special case of approximate range counting.
The goal is therefore to derive solutions that are comparable, in
their dependence on $n$, with those that solve emptiness (or reporting)
queries. As just noted, this has been accomplished for the case of
halfspaces. In this section we extend this technique to the general
semi-algebraic case.

The simplest solution is to adapt the technique of Aronov and
Har-Peled~\cite{AHP}, which uses a procedure for answering range
emptiness queries as a ``black box''. Specifically,
suppose we have a data structure, $\D(P')$, for any set $P'$ 
of $n'$ points, which can be constructed in $T(n')$ time, uses 
$S(n')$ storage, and can determine whether a query range
$\gamma \in \Gamma$ is empty in $Q(n')$ time.
Using such a black box, Aronov and Har-Peled show how to construct
a data structure for $n$ points using
$O((\delta^{\lambda-3} + \Sigma_{i=1}^{\lceil 1/\delta\rceil}
1/i^{\lambda-2})S(n)\log n)$ storage and 
$O((\delta^{\lambda-3} + \Sigma_{i=1}^{\lceil 1/\delta\rceil}
1/i^{\lambda-2})T(n)\log n)$ preprocessing,
where $\lambda \geq 1$ is some constant for which
$S(n/r) = O(S(n)/r^\lambda)$ and $T(n/r) = O(T(n)/r^\lambda)$, 
for any $r>1$.
Given a range $\gamma \in \Gamma$, the data structure of \cite{AHP}
returns, in $O(\delta^{-2}Q(n)\log n)$ time, an approximate count
$t_\gamma$, satisfying
$(1-\delta)|\gamma \cap P| \le t_\gamma \le |\gamma \cap P|$.

The intuition behind this approach is that a range $\gamma$,
containing $m$ points of $P$, is expected to contain $m r/n$ points
in a random sample from $P$ of size $r$, and no points in a sample 
of size smaller than $n/m$. The algorithm of \cite{AHP} then 
guesses the value of $a$ (up to a factor of $1+\delta$), sets $r$ to 
be an appropriate multiple of $n/m$, and draws many (specifically,
$O(\delta^{-2}\log n)$ random samples of size $r$. 
If $\gamma$ is empty (resp., nonempty)
for many of the samples then, with high probability, the guess 
for $m$ is too large (resp., too small). When we cannot decide 
either way, we are at the correct value of $m$ (up to a relative
error of $\delta$). The actual details of the search are somewhat more
contrived; see \cite{AHP} for those details.

Plugging our emptiness data structures into the machinery of
\cite{AHP}, we therefore obtain the following results. 
In all these applications we can take $\lambda=1$, so, in the
terminology used above, the overall data structure uses
$O(\delta^{-2}S(n)\log n)$ storage and
$O(\delta^{-2}T(n)\log n)$ preprocessing.

\begin{corollary}
Let $P$ be a set of $n$ points in the plane, and let $\alpha$, $\delta$
be given positive parameters. Then we can preprocess $P$ into a data
structure of size $O(\delta^{-2}n\log n)$, in time
$O(\delta^{-2}n^{1+\eps})$, for any $\eps>0$,
such that, for any $\alpha$-fat query triangle $\Delta$, we
can compute, in $O(\delta^{-2}n^\eps)$ time, for any $\eps>0$, 
an approximate count $t_\Delta$ satisfying
$(1-\delta)|\Delta \cap P| \le t_\Delta \le |\Delta \cap P|$.
\end{corollary}

\begin{corollary}
Let $P$ be a set of $n$ points in the plane, and let $\delta$
be a given positive parameter. Then we can preprocess $P$ into 
a data structure of size $O(\delta^{-2}n\log n)$, in time
$O(\delta^{-2}n^{1+\eps})$, for any $\eps>0$,
such that, for any line $\ell$, point $p$ on $\ell$ or above 
$\ell$, and distance $d$, we
can compute, in $O(\delta^{-2}n^\eps)$ time, for any $\eps>0$, 
an approximate count $t_{\ell,p,d}$ of the exact number 
$N_{\ell,p,d}$ of the points of $P$ which lie above $\ell$ 
and at distance at most $d$ from $p$, so that
$(1-\delta) N_{\ell,p,d} \le t_{\ell,p,d} \le N_{\ell,p,d}$.
\end{corollary}

\begin{corollary}
Let $P$ be a set of $n$ points in $\reals^3$, $\Gamma$ a 
collection of convex semi-algebraic ranges of constant 
description complexity, and $\delta$ a given positive parameter. 
Then we can preprocess $P$ into a data structure of size 
$O(\delta^{-2}n\log n)$, in time
$O(\delta^{-2}n^{1+\eps})$, for any $\eps>0$,
such that, for any query range $\gamma\in\Gamma$, we can
compute, in $O(\eps^{-2}n^{1/2+\eps}\log n)$ time, for any $\eps>0$,
an approximate count $t_\gamma$ of the number of points of $P$ 
outside $\gamma$, satisfying
$(1-\delta)|\gamma^c \cap P| \le t_\Delta \le |\gamma^c \cap P|$.
\end{corollary}

\begin{corollary}
Let $\B$ be a set of $n$ balls in $\reals^3$, and let
$\delta$ be a given positive parameter. 
Then we can preprocess $\B$ into a data structure of size 
$O(\delta^{-2}n\log n)$, in time
$O(\delta^{-2}n^{1+\eps})$, for any $\eps>0$,
such that, for any query ray $\rho$, we can 
compute, in $O(\eps^{-2}n^{2/3+\eps}\log n)$ time, for any $\eps>0$,
an approximate count $t_\rho$ of the exact number $N_\rho$ 
of the balls of $\B$ intersected by $\rho$, satisfying
$(1-\delta) N_\rho \le t_\rho \le N_\rho$.
\end{corollary}

\noindent{\bf Remark:}
Another approach to approximate range counting has been presented in
\cite{AHPS,ArS}, in which, rather than using range emptiness searching
as a black box, one modifies the partition tree of the range emptiness
data structure, and augments each of its inner nodes with a so-called 
{\em relative $(p,\eps)$ approximation} sets, which are then used to
obtain the approximate count of a range. This approach too can be
adapted to yield efficient approximate range counting algorithms for
semialgebraic ranges, with a slightly improved dependence of their
performance on $\delta$. We omit details of such an adaptation in this
paper.


\section{Conclusion}

In this paper we have presented a general approach to efficient range
emptiness searching with semi-algebraic ranges, and have applied it to
several specific emptiness searching and ray shooting problems. 
The present study resolves and overcomes
the technical problems encountered in our earlier study~\cite{ShSh2},
and presents more applications of the technique.

Clearly, there are many other applications of the new machinery, and
an obvious direction for further research is to ``dig them up''.
In each such problem, the main step would be to design a
good test set, with associated function $\zeta(\cdot)$ as small as
possible, using either the general recipe or an appropriate ad-hoc
analysis.
Many specific instances of this step are likely to generate interesting
(and often hard) combinatorial questions. For example, as already
mentioned earlier, we still do not know whether the complement of
the union of $n$ (congruent) cylinder in $\reals^3$ can be decomposed
into $O^*(n^2)$ elementary cells.

\end{document}